\documentclass[aps,superscriptaddress,preprintnumbers,showpacs]{revtex4}
\usepackage{amssymb}
\usepackage{graphicx}
\usepackage{dcolumn}
\usepackage{bm}

\begin{document}

\newcommand*{\PKU}{School of Physics and State Key Laboratory of Nuclear Physics and
Technology, Peking University, Beijing 100871}\affiliation{\PKU}
\newcommand*{\NTU}{Department of Physics and Center for Theoretical Sciences, National Taiwan University, Taipei 10617}\affiliation{\NTU}
\newcommand*{\chep}{Center for High Energy Physics, Peking University, Beijing 100871}\affiliation{\chep}

\title{Unified triminimal parametrizations of quark and lepton mixing matrices}

\author{Xiao-Gang He}\affiliation{\PKU}\affiliation{\NTU}\affiliation{\chep}
\author{Shi-Wen Li}\affiliation{\PKU}
\author{Bo-Qiang Ma}\email{mabq@phy.pku.edu.cn}\affiliation{\PKU}\affiliation{\chep}

\begin{abstract}
We present a detailed study on triminimal parametrizations of quark
and lepton mixing matrices with different basis matrices. We start
with a general discussion on the triminimal expansion of the mixing
matrix and on possible unified quark and lepton parametrization
using quark-lepton complementarity (QLC).  We then consider several
interesting basis matrices and compare the triminimal parametrizations with
the Wolfenstein-like parametrizations. The usual Wolfenstein
parametrization for quark mixing is a triminimal expansion around
the unit matrix as the basis. The corresponding QLC lepton mixing
matrix is a triminimal expansion around the bimaximal basis. Current
neutrino oscillation data show that the lepton mixing matrix is very
well represented by the tri-bimaximal mixing. It is natural to take
it as an expanding basis. The corresponding zeroth order basis for
quark mixing in this case makes the triminimal expansion converge
much faster than the usual Wolfenstein parametrization. The
triminimal expansion based on tri-bimaximal mixing can be converted
to the Wolfenstein-like parametrizations discussed in the
literature. We thus have a unified description between different
kinds of parametrizations for quark and lepton sectors: the standard
parametrizations, the Wolfenstein-like parametrizations, and the
triminimal parametrizations.

\end{abstract}


\maketitle

\section{Introduction}

Mixing between different generations of fermions in weak interaction
is one of the most interesting issues in particle physics. It is
characterized by an unitary matrix in the charged current
interaction of W-boson in the mass eigen-state of fermions. Quark
mixing is described by the
Cabibbo~\cite{cabibbo}-Kobayashi-Maskawa~\cite{km}(CKM) matrix
$V_{\rm{CKM}}$, and lepton mixing is described by the
Pontecorvo~\cite{pontecorvo}-Maki-Nakawaga-Sakata~\cite{mns} (PMNS)
matrix $U_{\rm{PMNS}}$ with
\begin{eqnarray}
L = -{g\over \sqrt{2}} \overline{U}_L \gamma^\mu V_{\rm CKM} D_L
W^+_\mu - {g\over \sqrt{2}} \overline{E}_L \gamma^\mu U_{\rm PMNS}
N_L W^-_\mu + H.C.\;,
\end{eqnarray}
where $U_L = (u_L,c_L,t_L,...)^T$, $D_L = (d_L,s_L,b_L,...)^T$, $E_L
= (e_L,\mu_L,\tau_L,...)^T$, and $N_L = (\nu_1,\nu_2,\nu_3,...)^T$
are the left-handed fermion generations. For n-generations, $V =
V_{\rm CKM}$ or $U_{\rm PMNS}$ is an $n\times n$ unitary matrix.

A commonly used form of mixing
matrix for three generations of fermions is given by~\cite{ck,yao},
\begin{eqnarray}
V = \left(
\begin{array}{ccc}
c_{12}c_{13} & s_{12}c_{13} & s_{13}e^{-i\delta}           \\
-s_{12}c_{23}-c_{12}s_{23}s_{13}e^{i\delta} &
c_{12}c_{23}-s_{12}s_{23}s_{13}e^{i\delta}  & s_{23}c_{13} \\
s_{12}s_{23}-c_{12}c_{23}s_{13}e^{i\delta}  &
-c_{12}s_{23}-s_{12}c_{23}s_{13}e^{i\delta} & c_{23}c_{13}
\end{array}
\right),\label{fp}
\end{eqnarray}
where $s_{ij}=\sin\theta_{ij}$ and $c_{ij}=\cos\theta_{ij}$ are the
mixing angles and $\delta$ is the CP violating phase. If neutrinos
are of Majorana type, for the PMNS matrix one should include an
additional diagonal matrix with two Majorana phases ${\rm
diag}(e^{i\alpha_1/2},e^{i\alpha_2/2},1)$ multiplied to the matrix
from right in the above. The two CP violating Majorana phases do not
affect neutrino oscillations, and we ignore these phases in our
discussions. To distinguish different CP violating phases, the phase
$\delta$ is sometimes called Dirac CP violating phase.

The above unitary matrix $V$ can be expressed as
\begin{eqnarray}
V = R_{23}(\theta_{23})U_\delta^\dag R_{13}(\theta_{13})U_\delta
R_{12}(\theta_{12})\;,\label{rrr}
\end{eqnarray}
with
\begin{eqnarray} R_{23}=\left(
\begin{array}{ccc}
1 & 0 & 0           \\
0 & c_{23} & s_{23} \\
0 & -s_{23} & c_{23}\\
\end{array}
\right)\;,\quad R_{13}=\left(
\begin{array}{ccc}
c_{13} & 0 & s_{13}           \\
0 & 1 & 0 \\
-s_{13} & 0 & c_{13}\\
\end{array}
\right)\;, \quad
R_{12} = \left(
\begin{array}{ccc}
c_{12}& s_{12} & 0           \\
-s_{12} & c_{12} & 0 \\
0 & 0 & 1\\
\end{array}
\right)\;,\label{ex}
\end{eqnarray}
and $U_{\delta}={\mathrm{diag}}(e^{i\delta/2}, 1, e^{-i\delta/2})$.
This way of parameterizing the mixing
provides a clear mathematic description of the mixing matrix with three angles
describing rotations in generation space and a phase describing
CP violation. In our later discussions, we will indicate the mixing angles with superscriptions
$Q$ and $L$ for quark and lepton sectors respectively
when specification is needed.

There are a lot of experimental data on the mixing patterns in both
the quark and lepton sectors. For quark mixing, the ranges of the
magnitudes of the CKM matrix elements have been very well determined
with~\cite{yao}
\begin{eqnarray} \left(
  \begin{array}{ccc}
    0.97419\pm0.00022             & 0.2257\pm0.0010    & 0.00359\pm0.00016               \\
    0.2256\pm0.0010               & 0.97334\pm0.00023  & 0.0415^{+0.0010}_{-0.0011}      \\
    0.00874^{+0.00026}_{-0.00037} & 0.0407\pm0.0010    & 0.999133^{+0.000044}_{-0.000043}
  \end{array} \right)\;.\label{vv}
\end{eqnarray}
From the above, we obtain the ranges for mixing angles $\theta^Q_{ij}$,
\begin{eqnarray}
\theta^Q_{12}=0.2277\pm0.0010,\quad
\theta^Q_{23}=0.0415^{+0.0010}_{-0.0011},\quad
\theta^Q_{13}=0.00359\pm0.00016. 
\end{eqnarray}
The CP violating phase has also been determined with
$\delta^Q\simeq\gamma^0=(66.7\pm 6.4)^\circ$~\cite{utfit}.

Considerable experimental data on lepton mixing have also been accumulated. The recent
global, $1\sigma(3\sigma)$, fit from neutrino oscillation data
gives~\cite{gonzalez},
\begin{eqnarray}
\theta_{12}^L=34.5^\circ\pm1.4^\circ(^{+4.8^\circ}_{-4.0^\circ}),\quad
\theta_{23}^L=42.3^{\circ+5.1^\circ}_{\,\,\,-3.3^\circ}(^{+11.3^\circ}_{-7.7^\circ}),\quad
\theta_{13}^L=0.0^{\circ+7.9^\circ}_{\,\,\,-0.0^\circ}(^{+12.9^\circ}_{-0.0^\circ}).
\label{langle}
\end{eqnarray}
At present there is no experimental data on the CP violating Dirac phase $\delta^L$ and Majorana phases $\alpha_i$.

When studying mixing, it is interesting to parameterize $V$
according to the hierarchical structure of the mixing to reveal more
physical information about the underlying theory. The Wolfenstein
parametrization for quarks is a famous example of this type, where
$V$ is parameterized as~\cite{wolfenstein}
\begin{eqnarray}
V_{CKM}=\left(
  \begin{array}{ccc}
    1-\frac{1}{2}\lambda^2   & \lambda                & A\lambda^3(\rho-i\eta) \\
    -\lambda                 & 1-\frac{1}{2}\lambda^2 & A\lambda^2             \\
    A\lambda^3(1-\rho-i\eta) & -A\lambda^2            & 1                      \\
  \end{array}
\right)+\mathcal{O}(\lambda^4), \label{wolfenstein}
\end{eqnarray}
with $\lambda=0.2257^{+0.0009}_{-0.0010}$,
$A=0.814^{+0.021}_{-0.022}$, $\rho(1-\lambda^2/2+\dots) =
0.135^{+0.031}_{-0.016}$, and $\eta(1-\lambda^2/2+\dots) =
0.349^{+0.015}_{-0.017}$~\cite{yao}. The parameter $\lambda$ serves
as a good indicator of hierarchy of the mixing phenomenon in quark
sector. Since when $\lambda$ goes to zero, the matrix $V_{\rm CKM}$
becomes a unit matrix, one can take the unit matrix as the zeroth
order basis in this perturbative expansion.

The Wolfenstein parametrization is therefore an expansion of $V_{\rm
CKM}$ around the unit matrix basis with $\lambda$ as the expanding
parameter. The connection to the usual three angle and one phase
parametrization can be identified as
\begin{eqnarray}
\lambda = s_{12}c_{13}\;,\quad A\lambda^2 = s_{23} c_{13}\;,\quad
A\lambda^3(\rho - i\eta) = s_{13} e^{-i\delta}\;.
\end{eqnarray}

One can then use this definition for the angles $\theta_{ij}$ and
the phase $\delta$ to make exact parametrization of $V_{\rm CKM}$
and expand it at an arbitrary power of $\lambda$. In this type of
parametrization, the choice of the parameters and
where to put them are arbitrary making the meaning of the parameters
subtle to some extent, for example the CP violating phase $\delta$
is not independent, i.e., it is determined by two parameters with,
$\tan\delta = \eta/\rho$.

It would be better to expand the mixing matrix $V$ with the
parameters kept small with clear physical meaning. One good choice
is that the expanding parameters also indicate mixing in generation
space. Then the procedure for finding a perturbative expanding
series is to identify the zeroth order mixing matrix $V_0$ and then
use three small mixing parameters and one CP violating phase to
expand the mixing matrix $V$. This is the triminimal
parametrization.

A good expansion is then judged by a good choice of $V_0$ such that
the expansion converges quickly. For the quark mixing, the unit
matrix is a reasonable zeroth order expansion, since the expanding
parameter $\lambda = 0.2257$ makes the convergence reasonably fast.
But the choice of unit matrix as the zeroth order matrix $V_0$ is
certainly not a good one for lepton sector, where it has been shown
experimentally that some of the off diagonal mixing elements are not
small. A direct copy of Wolfenstein parametrization for $U_{\rm
PMNS}$ is not suitable. In this situation, to incorporate
experimental information, it may be better to use the bimaximal
mixing matrix $U_{\rm{bi}}$~\cite{bi} or the tri-bimaximal mixing
matrix $U_{\rm{tri}}$~\cite{tri} as the zeroth order basis with
\begin{eqnarray}
U_{\rm{bi}}=\left(
              \begin{array}{ccc}
                1/\sqrt{2} & 1/\sqrt{2} & 0 \\
                -1/2       & 1/2        & 1/\sqrt{2} \\
                1/2        & -1/2       & 1/\sqrt{2} \\
              \end{array}
            \right),\quad
U_{\rm{tri}}=\left(
              \begin{array}{ccc}
              2/\sqrt{6}  & 1/\sqrt{3}  & 0          \\
              -1/\sqrt{6} & 1/\sqrt{3}  & 1/\sqrt{2} \\
              1/\sqrt{6}  & -1/\sqrt{3} & 1/\sqrt{2}
              \end{array}
            \right).
\end{eqnarray}
Although the bimaximal basis is not favored by present experimental
data, with corrections of order $\lambda$, it can accommodate
experimental data, the bimaximal basis is therefore a reasonable one
as good as the unit basis in quark mixing. The tri-bimaximal basis
is very close to the experimental mixing pattern. It is certainly a
good basis for lepton mixing expansion.

If one tries in a similar way to parameterize perturbation in
expansion in a Wolfenstein-like way, there are ambiguities in how to
choose the expanding parameters based on
$U_{\rm{bi}}$~\cite{bipara,bipara2} and
$U_{\rm{tri}}$~\cite{tripara,tripara2}. It would be desirable to
have a definitive way to make expansions. To this end, Pakvasa,
Rodejohann, and Weiler proposed the triminimal parametrization in
lepton sector~\cite{pakvasa}. The triminimal expansion of the quark
and lepton mixing pointed out a new way to parameterize the mixing
matrix with all angle parameters small, and with the CP violating
phase parameter free from other parameters. The parameters are
completely determined when the basis matrix is chosen.

The parametrizations of mixing for quark and lepton sectors, a
priori, seems unrelated. If there is a way to connect the seemingly
independent parametrizations of mixing in these two sectors, it
would gain more insights about fermion mixing. Indeed there is a
very nice way to make the connection via the so called quark-lepton
complementarity (QLC)~\cite{smirnov,qlc}. We find the QLC relations
very useful and will use it in our later discussions.

In this paper, with the help of QLC we present a detailed study of
parameterizing quark and lepton mixing matrices using the triminimal
parametrization technique with different basis matrices, and discuss
their relations with Wolfenstein-like parametrizations.

The organization of this paper is as follows: in Sec. II, we derive
the general expression of the triminimal parametrizations of the
mixing matrix. In Sec. III, we take the unit matrix as the basis of
the CKM matrix and the bimaximal matrix as the basis of the PMNS
matrix. We show that these two parametrizations are related by QLC.
In Sec. IV, we start from tri-bimaximal for lepton mixing, and then
use the QLC relations to determine the basis for the CKM matrix and
discuss some implications. In both Sec. III and IV, we also discuss
how Wolfenstein-like parametrizations can be obtained from
triminimal parametrizations. In Sec. V, we present our conclusions.

\section{The general expression of the triminimal parametrizations}

\subsection{The triminimal expansion}

The idea of the triminimal parametrization is to express a mixing
angle in the mixing matrices as the sum of a zeroth order angle $\theta^0$ and a small perturbation angle $\epsilon$ with
\begin{eqnarray}
\theta_{12}=\theta_{12}^0+\epsilon_{12},
\quad\theta_{23}=\theta_{23}^0+\epsilon_{23},\quad\theta_{13}=\theta_{13}^0+\epsilon_{13}.\label{theta}
\end{eqnarray}
With the deviations $\epsilon_{12,23,13}$, one can expand the
matrix elements in powers of $\epsilon_{12,23,13}$. We have
\begin{eqnarray}
V = R^\epsilon_{23}R^0_{23}U^\dagger_\delta
R^0_{13}R^\epsilon_{13}U_\delta R^0_{12}R^\epsilon_{12},
\end{eqnarray}
where $R^0_{ij} = R_{ij}(\theta^0_{ij})$ and $R^\epsilon_{ij} = R^\epsilon_{ij}(\epsilon_{ij})$.

The above can be written in a different form more suitable for
expansion in $\epsilon_{ij}$,
\begin{eqnarray}
V &=& (R^0_{23} + R^{0\prime}_{23}\sin\epsilon_{23} +
2R^{0\prime\prime}_{23}\sin^2\frac{\epsilon_{23}}{2})
U^\dagger_\delta (R^0_{13} + R^{0\prime}_{13}\sin\epsilon_{13} +
2R^{0\prime\prime}_{13}\sin^2\frac{\epsilon_{13}}{2})
U_\delta \nonumber\\
&\times& (R^0_{12} + R^{0\prime}_{12}\sin\epsilon_{12} +
2R^{0\prime\prime}_{12}\sin^2\frac{\epsilon_{12}}{2})\;,
\end{eqnarray}
where $R^{0\prime}_{ij} = \partial R^0_{ij}/\partial \theta^0_{ij}$
and $R^{0\prime\prime}_{ij} = \partial^2 R^0_{ij}/\partial
(\theta^0_{ij})^2$.

To second order in $\epsilon_{ij}$, the mixing matrix is given
by
\begin{eqnarray}
V&=&V_0+R^0_{23}U^\dagger_\delta R^0_{13}U_\delta R^{0\prime}_{12} \epsilon_{12}+R^{0\prime}_{23}U^\dagger_\delta
R^0_{13}U_\delta R^0_{12}\epsilon_{23}+R^0_{23}U^\dagger_\delta R^{0\prime}_{13}U_\delta R^0_{12}\epsilon_{13}\nonumber\\
&+&{1\over 2} R^0_{23}U^\dagger_\delta R^0_{13}U_\delta
R^{0\prime\prime}_{12} \epsilon^2_{12} +{1\over 2}
R^{0\prime\prime}_{23}U^\dagger_\delta R^0_{13}U_\delta R^{0}_{12}
\epsilon^2_{23}+{1\over 2} R^0_{23}U^\dagger_\delta
R^{0\prime\prime}_{13}U_\delta R^{0}_{12} \epsilon^2_{13}
\nonumber\\
&+&R^{0\prime}_{23}U^\dagger_\delta R^0_{13}U_\delta
R^{0\prime}_{12} \epsilon_{12}\epsilon_{23}+R^0_{23}U^\dagger_\delta
R^{0\prime}_{13}U_\delta R^{0\prime}_{12}
\epsilon_{12}\epsilon_{13}+ R^{0\prime}_{23}U^\dagger_\delta
R^{0\prime}_{13}U_\delta R^{0}_{12} \epsilon_{23}\epsilon_{13}
+\mathcal{O}(\epsilon_{ij}^3),\label{9v}
\end{eqnarray}
where $V_0 = R^0_{23}U^\dagger_\delta R^0_{13}U_\delta R^0_{12}$ is
the zeroth order expansion basis.

The Jarlskog parameter ~\cite{jarlskog} $J={\rm
Im}(V_{12}V_{23}V_{13}^\ast V_{32}^\ast)
={1\over8}\sin2\theta_{12}\sin2\theta_{23}\sin2\theta_{13}\cos\theta_{13}\sin\delta$
is phase-convention independent which makes it very important when
discussing CP violation. If CP violation is going to be discussed,
one should make sure that the expansion to a certain order, a
non-zero $J$ is obtained. The second order in $\epsilon_{ij}$ for
$J$ is given by
\begin{eqnarray}
J&=&J_0\Big(1+2\epsilon_{12}\cot2\theta^0_{12}+2\epsilon_{23}\cot2\theta_{23}^0
+\epsilon_{13}(2\cot2\theta^0_{13}-\tan\theta^0_{13})\nonumber\\
&-&2\epsilon_{12}^2-2\epsilon^2_{23}-\epsilon^2_{13}({7\over2}-\tan^2\theta^0_{13})
+4\epsilon_{12}\epsilon_{23}\cot2\theta^0_{12}\cot2\theta^0_{23}\nonumber\\
&+&2\epsilon_{12}\epsilon_{13}\cot2\theta^0_{12}(2\cot2\theta^0_{13}-\tan\theta^0_{13})
+2\epsilon_{23}\epsilon_{13}\cot2\theta^0_{23}(2\cot2\theta^0_{13}-\tan\theta^0_{13})\Big)+\mathcal{O}(\epsilon_{ij}^3),
\end{eqnarray}
where
$J_0={1\over8}\sin2\theta^0_{12}\sin2\theta^0_{23}\sin2\theta^0_{13}\cos\theta^0_{13}\sin\delta$.

Since $\theta_{13}$ in both quark and lepton sectors are very small,
a good choice for $\theta^0_{13}$ in Eq.~(\ref{theta}) is zero.
Then Eq.~(\ref{rrr}) is simplified to
\begin{eqnarray}
R_{23}(\theta^0_{23}+\epsilon_{23})U_\delta^\dag
R_{13}(\theta^0_{13}+\epsilon_{13})U_\delta
R_{12}(\theta^0_{12}+\epsilon_{12})=R_{23}(\theta^0_{23})R(\epsilon_{23})U_\delta^\dag
R_{13}(\epsilon_{13})U_\delta
R_{12}(\epsilon_{12})R_{12}(\theta^0_{12}).
\end{eqnarray}
Note that $R(\epsilon_{23})U_\delta^\dag R_{13}(\epsilon_{13})U_\delta
R_{12}(\epsilon_{12})$ is just Eq.~(\ref{fp}) with $\theta_{ij}$
replaced by $\epsilon_{ij}$.
$R_{23}(\theta^0_{23})R_{12}(\theta^0_{12})$ is the zeroth order
approximation of the mixing matrix $V_0$. In this case, Eq.~(\ref{9v}) is
simplified to be
\begin{eqnarray}
V&=&\left(
\begin{array}{ccc}
c^0_{12}c^0_{13}  & s^0_{12}c^0_{13}  & 0        \\
-s^0_{12}c^0_{23} & c^0_{12}c^0_{23}  & s^0_{23} \\
s^0_{12}s^0_{23}  & -c^0_{12}s^0_{23} & c^0_{23}
\end{array}
\right)+\epsilon_{12}\left(
\begin{array}{ccc}
-s^0_{12}         & c^0_{12}         & 0 \\
-c^0_{12}c^0_{23} &-s^0_{12}c^0_{23} & 0 \\
c^0_{12}s^0_{23}  & s^0_{12}s^0_{23} & 0
\end{array}
\right)+\epsilon_{23}\left(
\begin{array}{ccc}
0 & 0 & 0           \\
s^0_{12}s^0_{23} & -c^0_{12}s^0_{23} & c^0_{23}  \\
s^0_{12}c^0_{23} & -c^0_{12}c^0_{23} & -s^0_{23}
\end{array}
\right)\nonumber\\
&+&\epsilon_{13}\left(
\begin{array}{ccc}
0 & 0 & e^{-i\delta}  \\
-c^0_{12}s^0_{23}e^{i\delta}  & -s^0_{12}s^0_{23}e^{i\delta} & 0 \\
-c^0_{12}c^0_{23}e^{i\delta}  & -s^0_{12}c^0_{23}e^{i\delta} & 0
\end{array}
\right)+{1\over2}\epsilon^2_{12}\left(
\begin{array}{ccc}
-c^0_{12}         & -s^0_{12}         & 0 \\
s^0_{12}c^0_{23}  & -c^0_{12}c^0_{23} & 0 \\
-s^0_{12}s^0_{23} & c^0_{12}s^0_{23}  & 0
\end{array}
\right)+{1\over2}\epsilon^2_{23}\left(
\begin{array}{ccc}
0 & 0 & 0           \\
s^0_{12}c^0_{23}  & -c^0_{12}c^0_{23} & -s^0_{23} \\
-s^0_{12}s^0_{23} & c^0_{12}s^0_{23}  & -c^0_{23}
\end{array}
\right)\nonumber\\
&+&{1\over2}\epsilon^2_{13}\left(
\begin{array}{ccc}
-c^0_{12} & -s^0_{12} & 0         \\
0         & 0         & -s^0_{23} \\
0         & 0         & -c^0_{23}
\end{array}
\right)+\epsilon_{12}\epsilon_{23}\left(
\begin{array}{ccc}
0 & 0 & 0     \\
c^0_{12}s^0_{23} & s^0_{12}s^0_{23} & 0 \\
c^0_{12}c^0_{23} & s^0_{12}c^0_{23} & 0
\end{array}
\right)+\epsilon_{12}\epsilon_{13}\left(
\begin{array}{ccc}
0 & 0 & 0 \\
s^0_{12}s^0_{23}e^{i\delta}  & -c^0_{12}s^0_{23}e^{i\delta} & 0 \\
s^0_{12}c^0_{23}e^{i\delta}  & -c^0_{12}c^0_{23}e^{i\delta} & 0
\end{array}
\right)\nonumber\\
&+&\epsilon_{23}\epsilon_{13}\left(
\begin{array}{ccc}
0 & 0 & 0     \\
-c^0_{12}c^0_{23}e^{i\delta} & -s^0_{12}c^0_{23}e^{i\delta} & 0 \\
c^0_{12}s^0_{23}e^{i\delta}  & s^0_{12}s^0_{23}e^{i\delta}  & 0
\end{array}
\right)+\mathcal{O}(\epsilon_{ij}^3).\label{triminimalpara}
\end{eqnarray}

The Jarlskog parameter $J$ is then given by
\begin{eqnarray}
J = \left(\epsilon_{13}c_{12}^0c^0_{23}s^0_{12}s^0_{23}
+{1\over2}\epsilon_{12}\epsilon_{13}\sin2\theta^0_{23}\cos2\theta^0_{12}
+{1\over2}\epsilon_{23}\epsilon_{13}\sin2\theta^0_{12}\cos2\theta^0_{23}\right)\sin\delta
+ O(\epsilon^3_{ij}).
\end{eqnarray}

It is clear from above discussions that in general the expansion looks complicated. So
far the discussions are just a simple expansion in mathematics. A
good expansion should have the virtual being simple with fast
convergency. The choice of the zeroth order matrix $V_0$ which leads
to a simple expansion then reflects the physical insight of a
parametrization providing hints for underlying theory producing the
mixing.

\subsection{Unified parametrization through quark-lepton complementarity}

So far we have treated the parametrizations for quarks and leptons
separately. It would be interesting to find a unified
parametrization which connects these two seemingly unrelated
sectors. We find that the quark-lepton complementarity (QLC) can
provide a very useful guide for the unified treatment of mixing in
quark and lepton sectors. The QLC relations refer to
\begin{eqnarray}
\theta_{12}^Q+\theta_{12}^L=\frac{\pi}{4},
\quad\theta_{23}^Q+\theta_{23}^L=\frac{\pi}{4},
\quad\theta_{13}^Q\sim\theta_{13}^L\sim 0.\label{qlc}
\end{eqnarray}

If one writes
\begin{eqnarray}
&&\theta^Q_{12}=
\theta^{Q0}_{12}+\epsilon^Q_{12}\;,\quad\theta^Q_{23}=
\theta^{Q0}_{23} + \epsilon^Q_{23}\;,\quad
\theta^Q_{13}= \theta^{Q0}_{13}+\epsilon^Q_{13},\nonumber\\
&&\theta^L_{12}=
\theta^{L0}_{12}+\epsilon^L_{12}\;,\quad\theta^L_{23}=
\theta^{L0}_{23} + \epsilon^L_{23}\;,\quad\theta^L_{13}=
\theta^{L0}_{13}+\epsilon^L_{13},
\end{eqnarray}
a good choice of the zeroth order angles would be: $\theta^{Q0}_{12} + \theta^{L0}_{12} = \pi/4$,
$\theta^{Q0}_{23} + \theta^{L0}_{23} = \pi/4$, and $\theta^{Q0}_{13} = \theta^{L0}_{13} = 0$.
One has
\begin{eqnarray}
\epsilon^{Q}_{12} + \epsilon^{L}_{12} = 0\;, \quad\epsilon^Q_{23} +
\epsilon^L_{23} = 0\;,\quad\epsilon_{13}^Q\sim\epsilon_{13}^L\sim 0.
\end{eqnarray}
This procedure then leads to
\begin{eqnarray}
V_{\rm CKM} &=& R_{23}(\theta^{Q0}_{23})R_{23}(\epsilon_{23}^Q)U^\dagger_\delta R_{13}(\epsilon^Q_{13})U_\delta
R_{12}(\epsilon^Q_{12})R_{12}(\theta^0_{12})\;,\nonumber\\
U_{\rm PMNS} &=&
R_{23}({\pi\over4}-\theta^{Q0}_{23})R_{23}(\epsilon_{23}^L)U^\dagger_\delta
R_{13}(\epsilon^L_{13})
U_\delta R_{12}(\epsilon^L_{12})R_{12}({\pi\over4}-\theta^{Q0}_{12})\nonumber\\
&=&R_{23}({\pi\over4})R_{23}(-\theta^{Q0}_{23})R_{23}(-\epsilon_{23}^Q)U^\dagger_\delta
R_{13}(\epsilon^L_{13}) U_\delta
R_{12}(-\epsilon^Q_{12})R_{12}(-\theta^{Q0}_{12})R_{12}({\pi\over4})\;.
\end{eqnarray}

The corresponding Jarlskog parameters are also related,
\begin{eqnarray}
J^Q&=& {1\over 4}\sin2\theta^Q_{12}\sin2\theta^Q_{23} \sin\epsilon^Q_{13}\cos^2\epsilon^Q_{13}\sin\delta^Q\;,\nonumber\\
J^L &=& {1\over 4}\cos2\theta^Q_{12}\cos2\theta^Q_{23}
\sin\epsilon^L_{13}\cos^2\epsilon^L_{13}\sin\delta^L\;.
\end{eqnarray}

The above relations reveal an important difference in CP violation
in the quark and lepton sectors. Since phenomenologically,
$\theta^Q_{12}$ and $\theta^Q_{23}$ are both small, it is clear that
the absolute value of $J^L$ is much larger than the absolute value
of $J^Q$ if $\epsilon^L_{13}$ and $\delta^L$ are not much smaller
than $\epsilon^Q_{13}$ and $\delta^Q$.

In the following sections, we work out some specific choices of
$V_0$ and compare early studies with the triminimal parametrization
we are discussing here.

\section{Triminimal parametrizations of the CKM and PMNS matrices with unit and bimaximal basis matrices}

\subsection{The triminimal expansion}

Eq.~(\ref{vv}) shows that the off-diagonal elements in the CKM
matrix are all small compared with 1, it is natural to take the unit matrix as its
basis. Setting the expanding triminimal
parameters as
\begin{eqnarray}
\theta^Q_{12}=\epsilon^Q_{12},\quad\theta^Q_{23}=\epsilon^Q_{23},\quad\theta^Q_{13}=\epsilon^Q_{13},
\end{eqnarray}
and the CP-violating phase $\delta^Q$ for quark, we obtain the triminimal
parametrization of the CKM matrix as
\begin{eqnarray}
V_{\rm{CKM}}&=&R_{23}(\epsilon^Q_{23})U^\dag_\delta
R_{13}(\epsilon^Q_{13})U_\delta R_{12}(\epsilon^Q_{12}).
\end{eqnarray}

Numerically, we have
\begin{eqnarray}
\epsilon^Q_{12}=0.2277\pm0.0010,\quad
\epsilon^Q_{23}=0.0415^{+0.0010}_{-0.0011},\quad
\epsilon^Q_{13}=0.00359\pm0.00016.
\end{eqnarray}

To third order in $\epsilon^Q_{ij}$, we have
\begin{eqnarray}
V_{\rm CKM}&=&I
+ \left(\begin{array}{lll}   0 & \epsilon^Q_{12} & \epsilon^Q_{13}e^{-i\delta^Q} \\
                                           -\epsilon^Q_{12} & 0 & \epsilon^Q_{23} \\
                                            -\epsilon^Q_{13}e^{i\delta^Q} & -\epsilon^Q_{23} & 0 \\
                                           \end{array}
                                           \right) + \left(
                  \begin{array}{ccc}
                    -\frac{1}{2}(\epsilon^Q_{12})^2-\frac{1}{2}(\epsilon^Q_{13})^2  & 0 & 0 \\
                    -\epsilon^Q_{23}\epsilon^Q_{13}e^{i\delta^Q}
                    & -\frac{1}{2}(\epsilon^Q_{12})^2-\frac{1}{2}(\epsilon^Q_{23})^2 & 0 \\
                    \epsilon^Q_{12}\epsilon^Q_{23} & -\epsilon^Q_{12}\epsilon^Q_{13}e^{i\delta^Q}
                    & -\frac{1}{2}(\epsilon^Q_{23})^2-\frac{1}{2}(\epsilon^Q_{13})^2 \\
                  \end{array}
                \right)\nonumber\\
  &+&\left(
                  \begin{array}{ccc}
                    0 & -\frac{1}{6}(\epsilon^Q_{12})^3-\frac{1}{2}\epsilon^Q_{12}(\epsilon^Q_{13})^2
                    & -\frac{1}{6}(\epsilon^Q_{13})^3e^{-i\delta^Q} \\
                    \frac{1}{6}(\epsilon^Q_{12})^3+\frac{1}{2}\epsilon^Q_{12}(\epsilon^Q_{23})^2
                    & - \epsilon^Q_{12}\epsilon^Q_{23}\epsilon^Q_{13}e^{i\delta^Q}
                    & -\frac{1}{6} (\epsilon^Q_{23})^3 -\frac{1}{2}\epsilon^Q_{23}(\epsilon^Q_{13})^2 \\
                    \frac{1}{2} (\epsilon^Q_{12})^2\epsilon^Q_{13}e^{i\delta^Q}
                    +\frac{1}{2}(\epsilon^Q_{23})^2\epsilon^Q_{13}e^{i\delta^Q}
                    & \frac{1}{6} (\epsilon^Q_{23})^3+\frac{1}{2}(\epsilon^Q_{12})^2\epsilon^Q_{23}& 0 \\
                  \end{array}
                \right)
  +\mathcal{O}\left((\epsilon^Q_{ij})^4\right),
\end{eqnarray}
where $I$ denotes the $3\times3$ unit matrix.

In most cases, the approximation to second order in
$\epsilon^Q_{ij}$ is enough. Here we keep it to the third order
because that $\epsilon^Q_{13}$ is of order $(\epsilon^Q_{12})^3$,
for numerical consistency one should display terms of order
$(\epsilon^Q_{12})^3$ in the expansion if an expansion involves
$\epsilon^Q_{13}$. Also since $\epsilon^Q_{23}$ is of order
$(\epsilon^Q_{12})^2$, one should keep
$\epsilon^Q_{12}\epsilon^Q_{23}$ in the expansion. We then have a
numerical consistent expansion
\begin{eqnarray}
V_{\rm CKM}&=&I
+ \left(\begin{array}{ccc}   0 & \epsilon^Q_{12} & \epsilon^Q_{13}e^{-i\delta^Q} \\
                                           -\epsilon^Q_{12} & 0 & \epsilon^Q_{23} \\
                                            -\epsilon^Q_{13}e^{i\delta^Q} & -\epsilon^Q_{23} & 0 \\
                                           \end{array}
                                           \right)
  +\left(
                  \begin{array}{ccc}
                    -\frac{1}{2}(\epsilon^Q_{12})^2 & 0 & 0 \\
                    0 & -\frac{1}{2}(\epsilon^Q_{12})^2 & 0 \\
                   \epsilon^Q_{12}\epsilon^Q_{23} & 0 & 0 \\
                  \end{array}
                \right)\nonumber\\
  &+&\left(
                  \begin{array}{ccc}
                    0 & -\frac{1}{6}(\epsilon^Q_{12})^3 & 0 \\
                    \frac{1}{6}(\epsilon^Q_{12})^3 & 0 & 0 \\
                    0 & 0 & 0 \\
                  \end{array}
                \right)+\mathcal{O}\left((\epsilon^Q_{12})^4\right).\label{ckmw}
\end{eqnarray}

For lepton mixing matrix, if one use QLC as a guide, the corresponding expansion in trimnimal expansion would give
\begin{eqnarray}
U_{\rm{PMNS}}&=&R_{23}(\frac{\pi}{4})R_{23}(\epsilon^L_{23})U^\dag_\delta
R_{13}(\epsilon^L_{13})U_\delta
R_{12}(\epsilon^L_{12})R_{12}(\frac{\pi}{4}),
\end{eqnarray}
with $\epsilon^L_{12,23} = - \epsilon^Q_{12,23}$.

Numerically, the
$1\sigma(3\sigma)$ ranges of the triminimal parameters are
\begin{eqnarray}
(-0.25)-0.21<\epsilon^L_{12}<-0.16(-0.10),\quad
(-0.18)-0.10<\epsilon^L_{23}<0.04(0.15),\quad
(0.00)0.00<\epsilon^L_{13}<0.14(0.23).\label{un}
\end{eqnarray}
with the best fit values $\epsilon_{12}^L=-0.18$,
$\epsilon_{23}^L=-0.05$ and $\epsilon_{13}^L=0.00$.
Experimental data are consistent with QLC at 3$\sigma$ level. Should future
experimental data determine that QLC is inconsistent with data, one
will have to abandon QLC, but the expansion using bimaximal mixing
matrix $U_{bi}$ is still valid and can serve as one of the
expansions of the PMNS matrix.

To third order in $\epsilon^L_{ij}$, the expansion is
\begin{eqnarray}
U_{\rm PMNS}&=&U_{\rm bi}
+ \epsilon^L_{12}\left(\begin{array}{ccc}  -\frac{\sqrt{2}}{2} & \frac{\sqrt{2}}{2} & 0 \\
                                           -\frac{1}{2} & -\frac{1}{2} & 0 \\
                                           \frac{1}{2} & \frac{1}{2} & 0 \\
                                           \end{array}
                                           \right)
  +\epsilon^L_{23}\left(
                  \begin{array}{ccc}
                    0 & 0 & 0 \\
                    \frac{1}{2} & -\frac{1}{2} & \frac{\sqrt{2}}{2} \\
                    \frac{1}{2} & -\frac{1}{2} & -\frac{\sqrt{2}}{2} \\
                  \end{array}
                \right)
  +\epsilon^L_{13}\left(
                  \begin{array}{ccc}
                    0 & 0 & e^{-i\delta^L} \\
                    -\frac{1}{2}e^{i\delta^L} & -\frac{1}{2}e^{i\delta^L} & 0 \\
                    -\frac{1}{2}e^{i\delta^L} & -\frac{1}{2}e^{i\delta^L} & 0 \\
                  \end{array}
                \right)\nonumber\\
  &+&(\epsilon^L_{12})^2\left(
                  \begin{array}{ccc}
                    -\frac{\sqrt{2}}{4} & -\frac{\sqrt{2}}{4} & 0 \\
                    \frac{1}{4} & -\frac{1}{4} & 0 \\
                    -\frac{1}{4} & \frac{1}{4} & 0 \\
                  \end{array}
                \right)
  +(\epsilon^L_{23})^2\left(
                  \begin{array}{ccc}
                    0 & 0 & 0 \\
                    \frac{1}{4} & -\frac{1}{4} & -\frac{\sqrt{2}}{4} \\
                    -\frac{1}{4} & \frac{1}{4} & -\frac{\sqrt{2}}{4} \\
                  \end{array}
                \right)
  +(\epsilon^L_{13})^2\left(
                  \begin{array}{ccc}
                    -\frac{\sqrt{2}}{4} & -\frac{\sqrt{2}}{4} & 0 \\
                    0 & 0 & -\frac{\sqrt{2}}{4} \\
                    0 & 0 & -\frac{\sqrt{2}}{4} \\
                  \end{array}
                \right)\nonumber\\
  &+&\epsilon^L_{12}\epsilon^L_{23}\left(
                  \begin{array}{ccc}
                    0 & 0 & 0 \\
                    \frac{1}{2} & \frac{1}{2} & 0 \\
                    \frac{1}{2} & \frac{1}{2} & 0 \\
                  \end{array}
                \right)
  +\epsilon^L_{12}\epsilon^L_{13}e^{i\delta^L}\left(
                  \begin{array}{ccc}
                    0 & 0 & 0 \\
                    \frac{1}{2} & -\frac{1}{2} & 0 \\
                    \frac{1}{2} & -\frac{1}{2} & 0 \\
                  \end{array}
                \right)
  +\epsilon^L_{23}\epsilon^L_{13}e^{i\delta^L}\left(
                  \begin{array}{ccc}
                    0 & 0 & 0 \\
                    -\frac{1}{2} & -\frac{1}{2} & 0 \\
                    \frac{1}{2} & \frac{1}{2} & 0 \\
                  \end{array}
                \right)\nonumber\\
  &+&(\epsilon^L_{12})^3\left(
                  \begin{array}{ccc}
                    \frac{\sqrt{2}}{12} & -\frac{\sqrt{2}}{12} & 0 \\
                    \frac{1}{12} & \frac{1}{12} & 0 \\
                    -\frac{1}{12} & -\frac{1}{12} & 0 \\
                  \end{array}
                \right)
  +(\epsilon^L_{23})^3\left(
                  \begin{array}{ccc}
                    0 & 0 & 0 \\
                    -\frac{1}{12} & \frac{1}{12} & -\frac{\sqrt{2}}{12} \\
                    -\frac{1}{12} & \frac{1}{12} & \frac{\sqrt{2}}{12} \\
                  \end{array}
                \right)
  +(\epsilon^L_{13})^3\left(
                  \begin{array}{ccc}
                    0 & 0 & -\frac{1}{6}e^{-i\delta^L} \\
                    \frac{1}{12}e^{i\delta^L} & \frac{1}{12}e^{i\delta^L} & 0 \\
                    \frac{1}{12}e^{i\delta^L} & \frac{1}{12}e^{i\delta^L} & 0 \\
                  \end{array}
                \right)\nonumber\\
  &+&\epsilon^L_{12}(\epsilon^L_{23})^2\left(
                  \begin{array}{ccc}
                    0 & 0 & 0 \\
                    \frac{1}{4} & \frac{1}{4} & 0 \\
                    -\frac{1}{4} & -\frac{1}{4} & 0 \\
                  \end{array}
                \right)
  + (\epsilon^L_{12})^2\epsilon^L_{23}\left(
                  \begin{array}{ccc}
                    0 & 0 & 0 \\
                    -\frac{1}{4} & \frac{1}{4} & 0 \\
                    -\frac{1}{4} & \frac{1}{4} & 0 \\
                  \end{array}
                \right)
  +\epsilon^L_{12}(\epsilon^L_{13})^2\left(
                  \begin{array}{ccc}
                    \frac{\sqrt{2}}{4} & -\frac{\sqrt{2}}{4} & 0 \\
                    0 & 0 & 0 \\
                    0 & 0 & 0 \\
                  \end{array}
                \right)\nonumber\\
  &+&(\epsilon^L_{12})^2\epsilon^L_{13}e^{i\delta^L}\left(
                  \begin{array}{ccc}
                    0 & 0 & 0 \\
                    \frac{1}{4} & \frac{1}{4} & 0 \\
                    \frac{1}{4} & \frac{1}{4} & 0 \\
                  \end{array}
                \right)
  +\epsilon^L_{23}(\epsilon^L_{13})^2\left(
                  \begin{array}{ccc}
                    0 & 0 & 0 \\
                    0 & 0 & -\frac{\sqrt{2}}{4} \\
                    0 & 0 & \frac{\sqrt{2}}{4} \\
                  \end{array}
                \right)
  +(\epsilon^L_{23})^2\epsilon^L_{13}e^{i\delta^L}\left(
                  \begin{array}{ccc}
                    0 & 0 & 0 \\
                    \frac{1}{4} & \frac{1}{4} & 0 \\
                    \frac{1}{4} & \frac{1}{4} & 0 \\
                  \end{array}
                \right)\nonumber\\
  &+&\epsilon^L_{12}\epsilon^L_{23}\epsilon^L_{13}e^{i\delta^L}\left(
                  \begin{array}{ccc}
                    0 & 0 & 0 \\
                    \frac{1}{2} & -\frac{1}{2} & 0 \\
                    -\frac{1}{2} & \frac{1}{2} & 0 \\
                  \end{array}
                \right)+\mathcal{O}\left((\epsilon^L_{ij})^4\right).\label{exppan}
\end{eqnarray}

If $\epsilon^L_{ij}$ exhibit the same hierarchy as those in quark sector, that is,
$\epsilon^L_{23}\sim(\epsilon^L_{12})^2$, $\epsilon^L_{13}\sim(\epsilon^L_{12})^3$,
then to third order in $\epsilon^L_{12}$, we have
\begin{eqnarray}
U_{\rm PMNS}&=&U_{\rm bi}
+ \epsilon^L_{12}\left(\begin{array}{ccc}  -\frac{\sqrt{2}}{2} & \frac{\sqrt{2}}{2} & 0 \\
                                           -\frac{1}{2} & -\frac{1}{2} & 0 \\
                                           \frac{1}{2} & \frac{1}{2} & 0 \\
                                           \end{array}
                                           \right)
  +\epsilon^L_{23}\left(
                  \begin{array}{ccc}
                    0 & 0 & 0 \\
                    \frac{1}{2} & -\frac{1}{2} & \frac{\sqrt{2}}{2} \\
                    \frac{1}{2} & -\frac{1}{2} & -\frac{\sqrt{2}}{2} \\
                  \end{array}
                \right)
  +\epsilon^L_{13}\left(
                  \begin{array}{ccc}
                    0 & 0 & e^{-i\delta^L} \\
                    -\frac{1}{2}e^{i\delta^L} & -\frac{1}{2}e^{i\delta^L} & 0 \\
                    -\frac{1}{2}e^{i\delta^L} & -\frac{1}{2}e^{i\delta^L} & 0 \\
                  \end{array}
                \right)\nonumber\\
  &+&(\epsilon^L_{12})^2\left(
                  \begin{array}{ccc}
                    -\frac{\sqrt{2}}{4} & -\frac{\sqrt{2}}{4} & 0 \\
                    \frac{1}{4} & -\frac{1}{4} & 0 \\
                    -\frac{1}{4} & \frac{1}{4} & 0 \\
                  \end{array}
                \right)
  +\epsilon^L_{12}\epsilon^L_{23}\left(
                  \begin{array}{ccc}
                    0 & 0 & 0 \\
                    \frac{1}{2} & \frac{1}{2} & 0 \\
                    \frac{1}{2} & \frac{1}{2} & 0 \\
                  \end{array}
                \right)
 +(\epsilon^L_{12})^3\left(
                  \begin{array}{ccc}
                    \frac{\sqrt{2}}{12} & -\frac{\sqrt{2}}{12} & 0 \\
                    \frac{1}{12} & \frac{1}{12} & 0 \\
                    -\frac{1}{12} & -\frac{1}{12} & 0 \\
                  \end{array}
                \right)+\mathcal{O}\left((\epsilon^L_{12})^4\right).\label{pmnsw}
\end{eqnarray}

It is interesting to note that the basis of the PMNS
matrix is naturally the bimaximal mixing matrix $V_0 = U_{bi}$.

\subsection{Triminimal and Wolfenstein-like parametrizations}

We now compare the triminimal parametrization with Wolfenstein
parametrization for quark mixing. To obtain the Wolfenstein
parametrization, one may naively make the replacements of
$\epsilon^Q_{12}=\lambda$, $\epsilon^Q_{23}=A\lambda^2$, and
$\epsilon^Q_{13}e^{i\delta^Q}=A\lambda^3(\rho+i\eta)$, which lead to
\begin{eqnarray}
\epsilon^Q_{13}=A\lambda^3\sqrt{\rho^2+\eta^2},\quad
\epsilon^Q_{12}\epsilon^Q_{23}=A\lambda^{3}.
\end{eqnarray}

To third order in $\lambda$, one obtains
\begin{eqnarray}
V_{\rm CKM}=\left(
  \begin{array}{ccc}
    1-\frac{1}{2}\lambda^2   & \lambda                & A\lambda^3(\rho-i\eta) \\
    -\lambda                 & 1-\frac{1}{2}\lambda^2 & A\lambda^2             \\
    A\lambda^3(1-\rho-i\eta) & -A\lambda^2            & 1                      \\
  \end{array}\right)+\lambda^3\left(
                  \begin{array}{ccc}
                    0 & -\frac{1}{6} & 0 \\
                    \frac{1}{6} & 0 & 0 \\
                    0 & 0 & 0 \\
                  \end{array}
\right)+\mathcal{O}(\lambda^4).\label{wolfensteinadd}
\end{eqnarray}
The first term in the above is the Wolfenstein parametrization. But
there is an additional second term in the above equation. One may
wonder that the triminimal parametrization may not be equivalent to
the Wolfenstein parametrization. This is certainly not true. The
problem lies in the definition for $\lambda$. In the Wolfenstein
parametrization, $\lambda$ is defined through
\begin{eqnarray}
V_{us} = \cos\theta^Q_{13} \sin\theta^Q_{12}=\lambda.
\end{eqnarray}
But in the above naive identification of triminimal parametrization
with Wolfenstein parametrization, $\lambda$ is defined by
\begin{eqnarray}
\lambda = \epsilon^Q_{12}.
\end{eqnarray}
The above two definitions of $\lambda$'s are actually
different. Let us indicate temporarily by $\lambda' =
\epsilon^Q_{12}$, then to third order in $\lambda$ or $\lambda'$,
\begin{eqnarray}
\lambda = \sin\theta^Q_{12}=\epsilon^Q_{12}-{1\over6}(\epsilon^Q_{12})^3 = \lambda' - {1\over 6}(\lambda')^3.
\end{eqnarray}
Therefore $\lambda'$ is larger than $\lambda$ by a correct of order $\lambda^3$.

To correctly pass the triminimal parametrization to the usual Wolfenstein parametrization, one needs to define
\begin{eqnarray}
\epsilon^Q_{12}=\lambda+{1\over6}\lambda^3+\mathcal{O}(\lambda^4)\;,\quad\epsilon^Q_{23}=A
\lambda^2+\mathcal{O}(\lambda^4)\;,\quad\epsilon^Q_{13}e^{i\delta^Q}=A\lambda^3(\rho+i\eta)+\mathcal{O}(\lambda^4)\;.
\end{eqnarray}

With these definitions, the triminimal parametrization can be written in the same form
as the Wolfenstein parametrization with the same order of accuracy.

Now let us consider the mixing in the lepton sector seeking the
corresponding parametrization of the Wolfenstein-like
parametrization. In the quark sector, the largest triminimal
expansion parameter is $\epsilon^Q_{12} = \lambda$. One may start
with a similar set of expanding parameters with
the replacements $\epsilon^L_{12}=-\lambda$, $\epsilon^L_{23}=-A
\lambda^2$, and $\epsilon^L_{13}e^{i\delta^L}=A\lambda^3 (\zeta+i
\xi)$, then
\begin{eqnarray}
\epsilon^L_{13}=A\lambda^3\sqrt{\zeta^2+\xi^2},\quad
\epsilon^L_{12}\epsilon^L_{23}=A\lambda^{3}.
\end{eqnarray}

Comparing with the parametrization of {\it Case 1}
($\epsilon^L_{13}=|U_{e3}|\sim\lambda^3\sim(\epsilon^L_{12})^3$) in
Ref.~\cite{bipara2},
\begin{eqnarray}
U&=&U_{\rm bi}
+ \lambda\left(\begin{array}{ccc}  \frac{\sqrt{2}}{2} & -\frac{\sqrt{2}}{2} & 0 \\
                                           \frac{1}{2} & \frac{1}{2} & 0 \\
                                           -\frac{1}{2} & -\frac{1}{2} & 0 \\
                                           \end{array}
                                           \right)
  +\lambda^2\left(
                  \begin{array}{ccc}
                    -\frac{\sqrt{2}}{4} & -\frac{\sqrt{2}}{4} & 0 \\
                    -\frac{1}{2}(A-\frac{1}{2}) & \frac{1}{2}(A-\frac{1}{2}) & -\frac{\sqrt{2}}{2}A \\
                    -\frac{1}{2}(A+\frac{1}{2}) & \frac{1}{2}(A+\frac{1}{2}) & \frac{\sqrt{2}}{2}A \\
                  \end{array}
                \right)\nonumber\\
  &+&\lambda^3\left(
                  \begin{array}{ccc}
                    0 & 0 & A(\zeta-i\xi) \\
                    \frac{1}{2}A(1-\zeta-i\xi) & \frac{1}{2}A(1-\zeta-i\xi) & 0 \\
                    \frac{1}{2}A(1-\zeta-i\xi) & \frac{1}{2}A(1-\zeta-i\xi) & 0 \\
                  \end{array}
                \right)+\mathcal{O}(\lambda^4),\label{case1}
\end{eqnarray}
there is an additional term result from Eq.~(\ref{pmnsw}) given by
\begin{eqnarray}
\lambda^3\left(
                  \begin{array}{ccc}
                    -\frac{\sqrt{2}}{12} & \frac{\sqrt{2}}{12} & 0 \\
                    -\frac{1}{12} & -\frac{1}{12} & 0 \\
                    \frac{1}{12} & \frac{1}{12} & 0 \\
                  \end{array}
                \right).\label{add}
\end{eqnarray}

This difference is again caused by different definition of
$\lambda$. In Ref.~\cite{bipara2} it is defined as (to order
$\lambda^3$)
\begin{eqnarray}
\sin\theta^L_{12}=\frac{\sqrt{2}}{2}(1-\lambda-{1\over2}\lambda^2).
\end{eqnarray}
However, in our triminimal parametrization $\lambda =
-\epsilon^L_{12}$ is defined as (to order $(\epsilon^L_{12})^3$)
\begin{eqnarray}
\sin\theta^L_{12}
=\frac{\sqrt{2}}{2}\left(1+\epsilon^L_{12}-{1\over2}(\epsilon^L_{12})^2-{1\over6}(\epsilon^L_{12})^3\right).
\end{eqnarray}

If one replaces $\epsilon^L_{12}$ by
$-\lambda-{1\over6}\lambda^3+\mathcal{O}(\lambda^4)$,
$\epsilon^L_{23}$ by $-A\lambda^2+\mathcal{O}(\lambda^4)$, and
$\epsilon^L_{13}e^{i\delta^L}$ by
$A\lambda^3(\zeta+i\xi)+\mathcal{O}(\lambda^4)$, the resulting
mixing matrix from triminimal expansion give the same one as in
eq.(\ref{case1}). The two Wolfenstein-like parametrizations in
Ref.~\cite{bipara2} and the one resulting from triminimal
parametrization are equivalent.

If $|U_{e3}|$ is not very small, but with
$\epsilon^L_{13}\sim(\epsilon^L_{12})^2$, which is the {\it Case 2}
discussed in Ref.~\cite{bipara2}, then to third order in
$\epsilon^L_{12}$, we will need to keep in Eq.~(\ref{pmnsw}) the
$\epsilon^L_{12}\epsilon^L_{13}$ term in Eq.~(\ref{exppan}). In the
parametrization of {\it Case 2}
($\epsilon^L_{13}=|U_{e3}|\sim\lambda^2\sim(\epsilon^L_{12})^2$) in
Ref.~\cite{bipara2}, the expansion of the PMNS matrix is

\begin{eqnarray}
U&=&U_{\rm bi}
+ \lambda\left(\begin{array}{ccc}  \frac{\sqrt{2}}{2} & -\frac{\sqrt{2}}{2} & 0 \\
                                           \frac{1}{2} & \frac{1}{2} & 0 \\
                                           -\frac{1}{2} & -\frac{1}{2} & 0 \\
                                           \end{array}
                                           \right)
  +\lambda^2\left(
                  \begin{array}{ccc}
                    -\frac{\sqrt{2}}{4} & -\frac{\sqrt{2}}{4} & A(\zeta'-i\xi') \\
                    \frac{1}{2}\left(\frac{1}{2}-A(1+\zeta'+i\xi')\right)
                    & -\frac{1}{2}\left(\frac{1}{2}-A(1-\zeta'-i\xi')\right) & -\frac{\sqrt{2}}{2}A \\
                    -\frac{1}{2}\left(\frac{1}{2}+A(1+\zeta'+i\xi')\right)
                    & \frac{1}{2}\left(\frac{1}{2}+A(1-\zeta'-i\xi')\right) & \frac{\sqrt{2}}{2}A \\
                  \end{array}
                \right)\nonumber\\
  &+&\mathcal{O}(\lambda^3).\label{case2}
\end{eqnarray}
Making the replacements $\epsilon^L_{12}=-\lambda$,
$\epsilon^L_{23}=-A \lambda^2$, and
$\epsilon^L_{13}e^{i\delta^L}=A\lambda^2 (\zeta'+i \xi')$ which
means
\begin{eqnarray}
\epsilon^L_{13}=A\lambda^2\sqrt{\zeta'^2+\xi'^2},\quad
\epsilon^L_{12}\epsilon^L_{23}=A\lambda^{3},
\end{eqnarray}
then Eq.~(\ref{pmnsw}) is translated into Eq.~(\ref{case2}). Since the
expansion of Eq.~(\ref{case2}) is an approximation to
$\mathcal{O}(\lambda^2)$, the terms of $(\epsilon^L_{12})^3$,
$\epsilon^L_{12}\epsilon^L_{23}$ and
$\epsilon^L_{12}\epsilon^L_{13}$ in Eq.~(\ref{pmnsw}) do not need to
taken into account as $\mathcal{O}(\lambda^3)$.

To the lowest order the Jarlskog parameter are given by
\begin{eqnarray}
&&\mathcal{J}^Q={\rm{Im}}(V_{us}V_{cb}V_{ub}^\ast V_{cs}^\ast)
=\epsilon^Q_{12}\epsilon^Q_{23}\epsilon^Q_{13}\sin{\delta^Q},\nonumber\\
&&\mathcal{J}^L={\rm{Im}}(U_{e2}U_{\mu3}U_{e3}^\ast
U_{\mu2}^\ast)={1\over4}\epsilon^L_{13}\sin{\delta^L}.\label{jql}
\end{eqnarray}
The two results are very simple and may be useful in experimental
analysis. The expression of $\mathcal{J}^L$ is consistent with that
in Ref.~\cite{bipara2} to lowest order in $\lambda$.

We end this section with the conclusions that one can derive the
Wolfenstein parametrization, and the Wolfenstein-like
parametrization of the PMNS matrix in Ref.~\cite{bipara2} from the
triminimal parametrization in basis of the unit matrix and the
bimaximal matrix. The parametrization of the PMNS matrix in
Ref.~\cite{bipara2} is unified with the Wolfenstein parametrization
with the same parameters $\lambda$ and $A$ using the QLC relations.
The QLC relates unit matrix basis in quark sector to the bimaximal
basis in lepton sector.

\section{Triminimal parametrizations of the CKM and PMNS matrices in tri-bimaximal pattern}

\subsection{The triminimal expansion}

It is interesting to note that starting with unit matrix for $V_0$
for quark mixing expansion, QLC leads to bimaximal mixing matrix for
$V_0$ for lepton sector. Present data, however, indicate that the
mixing in the lepton sector is closer to the tri-bimaximal mixing.
One would have a much faster convergence if an expansion for lepton
sector starts with $V_0$ which has a tri-bimaximal mixing form.

In Ref.~\cite{pakvasa}, the triminimal parametrization of the PMNS
matrix is obtained as
\begin{eqnarray}
U_{\mathrm{PMNS}} &=&U_{\rm tri}
+ \epsilon^L_{12}\left(\begin{array}{ccc}  -\frac{1}{\sqrt{3}} & \frac{2}{\sqrt{6}}  & 0 \\
                                           -\frac{1}{\sqrt{3}} & -\frac{1}{\sqrt{6}} & 0 \\
                                           \frac{1}{\sqrt{3}}  & \frac{1}{\sqrt{6}}  & 0 \\
                                           \end{array}
                                           \right)
  +\epsilon^L_{23}\left(
                  \begin{array}{ccc}
                    0 & 0 & 0 \\
                    \frac{1}{\sqrt{6}} & -\frac{1}{\sqrt{3}} & \frac{1}{\sqrt{2}} \\
                    \frac{1}{\sqrt{6}} & -\frac{1}{\sqrt{3}} & -\frac{1}{\sqrt{2}} \\
                  \end{array}
                \right)
  +\epsilon^L_{13}\left(
                  \begin{array}{ccc}
                    0 & 0 & e^{-i\delta^L} \\
                    -\frac{1}{\sqrt{3}}e^{i\delta^L} & -\frac{1}{\sqrt{6}}e^{i\delta^L} & 0 \\
                    -\frac{1}{\sqrt{3}}e^{i\delta^L} & -\frac{1}{\sqrt{6}}e^{i\delta^L} & 0 \\
                  \end{array}
                \right)\nonumber\\
  &+&(\epsilon^L_{12})^2\left(
                  \begin{array}{ccc}
                    -\frac{1}{\sqrt{6}} & -\frac{1}{2\sqrt{3}} & 0 \\
                    \frac{1}{2\sqrt{6}} & -\frac{1}{2\sqrt{3}} & 0 \\
                    -\frac{1}{2\sqrt{6}} & \frac{1}{2\sqrt{3}} & 0 \\
                  \end{array}
                \right)
  +(\epsilon^L_{23})^2\left(
                  \begin{array}{ccc}
                    0 & 0 & 0 \\
                    \frac{1}{2\sqrt{6}} & -\frac{1}{2\sqrt{3}} & -\frac{1}{2\sqrt{2}} \\
                    -\frac{1}{2\sqrt{6}} & \frac{1}{2\sqrt{3}} & -\frac{1}{2\sqrt{2}} \\
                  \end{array}
                \right)
  +(\epsilon^L_{13})^2\left(
                  \begin{array}{ccc}
                    -\frac{1}{\sqrt{6}} & -\frac{1}{2\sqrt{3}} & 0 \\
                    0 & 0 & -\frac{1}{2\sqrt{2}} \\
                    0 & 0 & -\frac{1}{2\sqrt{2}} \\
                  \end{array}
                \right)\nonumber\\
  &+&\epsilon^L_{12}\epsilon^L_{23}\left(
                  \begin{array}{ccc}
                    0 & 0 & 0 \\
                    \frac{1}{\sqrt{3}} & \frac{1}{\sqrt{6}} & 0 \\
                    \frac{1}{\sqrt{3}} & \frac{1}{\sqrt{6}} & 0 \\
                  \end{array}
                \right)
  +\epsilon^L_{12}\epsilon^L_{13}e^{i\delta^L}\left(
                  \begin{array}{ccc}
                    0 & 0 & 0 \\
                    \frac{1}{\sqrt{6}} & -\frac{1}{\sqrt{3}} & 0 \\
                    \frac{1}{\sqrt{6}} & -\frac{1}{\sqrt{3}} & 0 \\
                  \end{array}
                \right)
  +\epsilon^L_{23}\epsilon^L_{13}e^{i\delta^L}\left(
                  \begin{array}{ccc}
                    0 & 0 & 0 \\
                    -\frac{1}{\sqrt{3}} & -\frac{1}{\sqrt{6}} & 0 \\
                    \frac{1}{\sqrt{3}} & \frac{1}{\sqrt{6}} & 0 \\
                  \end{array}
                \right),\label{prw}
\end{eqnarray}
with $(-0.08)-0.04<\epsilon^L_{12}<0.01(-0.07)$, $\epsilon^L_{23}$,
$\epsilon^L_{13}$, $\delta^L$ are the same parameters to the
parametrization of the PMNS matrix in Sec. III. This set of
expansion parameters is certainly better than the one in the
previous section if convergency is the criteria for the expansion.

With QLC, this leads to a very different $V_0$ for quark mixing than
the Wolfenstein parametrization. In Ref.~\cite{he}, we have derived
the triminimal parametrization of the CKM matrix
\begin{eqnarray}
V_{\mathrm{CKM}} &=&V_0
+ \epsilon^Q_{12}\left(\begin{array}{ccc}   -\frac{\sqrt{2}-1}{\sqrt{6}} &  \frac{\sqrt{2}+1}{\sqrt{6}} & 0 \\
                                           -\frac{\sqrt{2}+1}{\sqrt{6}} & -\frac{\sqrt{2}-1}{\sqrt{6}} & 0 \\
                                           0 & 0 & 0 \\
                                           \end{array}
                                           \right)
  +\epsilon^Q_{23}\left(
                  \begin{array}{ccc}
                    0 & 0 & 0 \\
                    0 & 0 & 1 \\
                    \frac{\sqrt{2}-1}{\sqrt{6}} & -\frac{\sqrt{2}+1}{\sqrt{6}} & 0 \\
                  \end{array}
                \right)
  +\epsilon^Q_{13}\left(
                  \begin{array}{ccc}
                    0 & 0 & e^{-i\delta} \\
                    0 & 0 & 0 \\
                    -\frac{\sqrt{2}+1}{\sqrt{6}}e^{i\delta} & -\frac{\sqrt{2}-1}{\sqrt{6}}e^{i\delta} & 0 \\
                  \end{array}
                \right)\nonumber\\
  &+&(\epsilon^Q_{12})^2\left(
                  \begin{array}{ccc}
                    -\frac{\sqrt{2}+1}{2\sqrt{6}} & -\frac{\sqrt{2}-1}{2\sqrt{6}} & 0 \\
                    \frac{\sqrt{2}-1}{2\sqrt{6}} & -\frac{\sqrt{2}+1}{2\sqrt{6}} & 0 \\
                    0 & 0 & 0 \\
                  \end{array}
                \right)
  +(\epsilon^Q_{23})^2\left(
                  \begin{array}{ccc}
                    0 & 0 & 0 \\
                    \frac{\sqrt{2}-1}{2\sqrt{6}} & -\frac{\sqrt{2}+1}{2\sqrt{6}} & 0 \\
                    0 & 0 & -\frac{1}{2} \\
                  \end{array}
                \right)
  +(\epsilon^Q_{13})^2\left(
                  \begin{array}{ccc}
                    -\frac{\sqrt{2}+1}{2\sqrt{6}} & -\frac{\sqrt{2}-1}{2\sqrt{6}} & 0 \\
                    0 & 0 & 0 \\
                    0 & 0 & -\frac{1}{2} \\
                  \end{array}
                \right)\nonumber\\
  &+&\epsilon^Q_{12}\epsilon^Q_{23}\left(
                  \begin{array}{ccc}
                    0 & 0 & 0 \\
                    0 & 0 & 0 \\
                    \frac{\sqrt{2}+1}{\sqrt{6}} & \frac{\sqrt{2}-1}{\sqrt{6}} & 0 \\
                  \end{array}
                \right)
  +\epsilon^Q_{12}\epsilon^Q_{13}e^{i\delta}\left(
                  \begin{array}{ccc}
                    0 & 0 & 0 \\
                    0 & 0 & 0 \\
                    \frac{\sqrt{2}-1}{\sqrt{6}} & -\frac{\sqrt{2}+1}{\sqrt{6}} & 0 \\
                  \end{array}
                \right)
  +\epsilon^Q_{23}\epsilon^Q_{13}e^{i\delta}\left(
                  \begin{array}{ccc}
                    0 & 0 & 0 \\
                    -\frac{\sqrt{2}+1}{\sqrt{6}} & -\frac{\sqrt{2}-1}{\sqrt{6}} & 0 \\
                    0 & 0 & 0 \\
                  \end{array}
                \right), \label{ve}
\end{eqnarray}
with $\epsilon^Q_{12}=0.0577\pm0.0010$, $\epsilon^Q_{23}$,
$\epsilon^Q_{13}$, $\delta^Q$ are the same parameters to the
parametrization of the CKM matrix in Sec. III, and
\begin{eqnarray}
    V_0=\left(
        \begin{array}{ccc}
            \frac{\sqrt{2}+1}{\sqrt{6}} & \frac{\sqrt{2}-1}{\sqrt{6}} & 0 \\
           -\frac{\sqrt{2}-1}{\sqrt{6}} & \frac{\sqrt{2}+1}{\sqrt{6}} & 0 \\
            0                           & 0                           & 1
        \end{array}
        \right).\label{vb}
\end{eqnarray}

\subsection{The triminimal and Wolfenstein-like parametrizations}

In Ref.~\cite{tripara2}, the Wolfenstein-like parametrization of the
PMNS matrix is also derived in two cases depending on the value of
$|U_{e3}|$. Let us discuss the relation between the two
Wolfenstein-like parametrizations and the triminimal parametrization
here.

If $|U_{e3}|$ is of order $\lambda^2$, {\it Case 1} in Ref.~\cite{tripara2}, after making the replacements of
$\epsilon^L_{12}=-\sqrt{6}(\sqrt{2}-1)\lambda-3(\sqrt{2}-1)^4\lambda^2+\mathcal{O}(\lambda^3)$,
$\epsilon^L_{23}=-A \lambda+\mathcal{O}(\lambda^3)$, and
$\epsilon^L_{13}e^{i\delta^L}=A\lambda^2(\zeta+i\xi)+\mathcal{O}(\lambda^3)=A\lambda^2z+\mathcal{O}(\lambda^3)$,
we have
\begin{eqnarray}
\epsilon^L_{13}=A\lambda^2\sqrt{\zeta^2+\xi^2}+\mathcal{O}(\lambda^3),
\quad\epsilon^L_{12}\epsilon^L_{23}=\sqrt{6}(\sqrt{2}-1)A\lambda^2+\mathcal{O}(\lambda^3).
\end{eqnarray}
We obtain the parametrization of {\it Case 1} of the PMNS matrix in
Ref.~\cite{tripara2} from Eq.~(\ref{prw})
\begin{eqnarray}
U&=&U_{\rm tri}+\lambda \left(
  \begin{array}{ccc}
    2-\sqrt{2}                        & -(2\sqrt{2}-2)                    & 0                    \\
    2-\sqrt{2}-\frac{1}{\sqrt{6}}A    & \sqrt{2}-1+\frac{1}{\sqrt{3}}A    & -\frac{\sqrt{2}}{2}A \\
    -(2-\sqrt{2})-\frac{1}{\sqrt{6}}A & -(\sqrt{2}-1)+\frac{1}{\sqrt{3}}A & \frac{\sqrt{2}}{2}A
  \end{array}
\right)\nonumber\\
&+&\lambda^2\left(
  \begin{array}{ccc}
    -(15\sqrt{6}-21\sqrt{3}) & -(15\sqrt{6}-21\sqrt{3}) & z^\ast A \\
    15\sqrt{3}-\frac{63}{\sqrt{6}}+\left(2-\sqrt{2}-\frac{z}{\sqrt{3}}\right)A+\frac{1}{2\sqrt{6}}A^2
    & -\left(15\sqrt{3}-\frac{63}{\sqrt{6}}\right)+\left(\sqrt{2}-1-\frac{z}{\sqrt{6}}\right)A
    -\frac{1}{2\sqrt{3}}A^2  & -\frac{\sqrt{2}}{4}A^2 \\
    -\left(15\sqrt{3}-\frac{63}{\sqrt{6}}\right)+\left(2-\sqrt{2}-\frac{z}{\sqrt{3}}\right)A-\frac{1}{2\sqrt{6}}A^2
    & 15\sqrt{3}-\frac{63}{\sqrt{6}}+\left(\sqrt{2}-1-\frac{z}{\sqrt{6}}\right)A
    +\frac{1}{2\sqrt{3}}A^2  & -\frac{\sqrt{2}}{4}A^2
  \end{array}
\right)\nonumber\\
&+&\mathcal{O}(\lambda^3).\label{ue2}
\end{eqnarray}

If $|U_{e3}|$ is of order $\lambda$, {\it Case 2} in Ref.~\cite{tripara2}, making the replacements of
$\epsilon^L_{12}=-\sqrt{6}(\sqrt{2}-1)\lambda-3(\sqrt{2}-1)^4\lambda^2+\mathcal{O}(\lambda^3)$,
$\epsilon^L_{23}=-A \lambda+\mathcal{O}(\lambda^3)$, and
$\epsilon^L_{13}e^{i\delta^L}=A\lambda(\zeta'+i\xi')+\mathcal{O}(\lambda^3)=A\lambda
z'+\mathcal{O}(\lambda^3)$, we have
\begin{eqnarray}
\epsilon^L_{13}=A\lambda\sqrt{\zeta'^2+\xi'^2}+\mathcal{O}(\lambda^3),
\quad\epsilon^L_{12}\epsilon^L_{23}=\sqrt{6}(\sqrt{2}-1)A\lambda^2+\mathcal{O}(\lambda^3).
\end{eqnarray}
We obtain the parametrization of {\it Case 2} of the PMNS matrix
in Ref.~\cite{tripara2} from Eq.~(\ref{prw})
\begin{eqnarray}
U&=&U_{\rm tri}+\lambda \left(
  \begin{array}{ccc}
    2-\sqrt{2} & -(2\sqrt{2}-2) & z'^\ast A \\
    2-\sqrt{2}-\left(\frac{1}{\sqrt{6}}+\frac{z'}{\sqrt{3}}\right)A
    & \sqrt{2}-1+\left(\frac{1}{\sqrt{3}}-\frac{z'}{\sqrt{6}}\right)A    & -\frac{\sqrt{2}}{2}A \\
    -(2-\sqrt{2})-\left(\frac{1}{\sqrt{6}}+\frac{z'}{\sqrt{3}}\right)A
    & -(\sqrt{2}-1)+\left(\frac{1}{\sqrt{3}}-\frac{z'}{\sqrt{6}}\right)A & \frac{\sqrt{2}}{2}A
  \end{array}
\right)\nonumber\\
&+&\lambda^2\left(
  \begin{array}{ccc}
    -(15\sqrt{6}-21\sqrt{3})-\frac{\sqrt{6}}{6}|z'|^2A^2 & -(15\sqrt{6}-21\sqrt{3})-\frac{\sqrt{3}}{6}|z'|^2A^2 & 0  \\
    15\sqrt{3}-\frac{63}{\sqrt{6}}+mA+pA^2               & -\left(15\sqrt{3}-\frac{63}{\sqrt{6}}\right)+nA-qA^2
    & -\frac{\sqrt{2}}{4}(1+|z'|^2)A^2 \\
    -\left(15\sqrt{3}-\frac{63}{\sqrt{6}}\right)+mA-pA^2 & 15\sqrt{3}-\frac{63}{\sqrt{6}}+nA+qA^2
    & -\frac{\sqrt{2}}{4}(1+|z'|^2)A^2
  \end{array}
\right)+\mathcal{O}(\lambda^3),\label{ue1}
\end{eqnarray}
where
\begin{eqnarray}
m=2-\sqrt{2}-\frac{z'}{\sqrt{2}+1},\quad
n=\sqrt{2}-1+\frac{2z'}{2+\sqrt{2}},\quad
p=\frac{1}{2\sqrt{6}}+\frac{z'}{\sqrt{3}},\quad
q=\frac{1}{2\sqrt{3}}-\frac{z'}{\sqrt{6}}.\nonumber
\end{eqnarray}

We have seen in both the last section and this section, the
Wolfenstein-like parametrization of the PMNS matrix depends on how
$U_{e3}$ is parameterized. However, the triminimal parametrization
is independent of how $U_{e3}$ is parameterized making it easier to
work with. One can obtain the Wolfenstein-like
parametrization from the triminimal parametrization when appropriate
definitions of the parameters are used.

For mixing in the quark sector, when taking the following definitions of the small parameters as,
$\epsilon^Q_{12}=\sqrt{6}(\sqrt{2}-1)\lambda+3(\sqrt{2}-1)^4\lambda^2+\mathcal{O}(\lambda^3)$,
$\epsilon^Q_{23}=A\lambda+\mathcal{O}(\lambda^3)$, and
$\epsilon^Q_{13}e^{i\delta^L}=A\lambda^2(\rho+i\eta)+\mathcal{O}(\lambda^3)$,
we have
\begin{eqnarray}
\epsilon^Q_{13}=A\lambda^2\sqrt{\rho^2+\eta^2}+\mathcal{O}(\lambda^3),
\quad\epsilon^Q_{12}\epsilon^Q_{23}=\sqrt{6}(\sqrt{2}-1)A\lambda^2+\mathcal{O}(\lambda^3).
\end{eqnarray}
Then we get the parametrization of the CKM matrix in
Ref.~\cite{tripara2}
\begin{eqnarray}
    V&=&\left(
        \begin{array}{ccc}
            \frac{\sqrt{2}+1}{\sqrt{6}} & \frac{\sqrt{2}-1}{\sqrt{6}} & 0 \\
           -\frac{\sqrt{2}-1}{\sqrt{6}} & \frac{\sqrt{2}+1}{\sqrt{6}} & 0 \\
            0                           & 0                           & 1
        \end{array}
        \right)+\lambda\left(
        \begin{array}{ccc}
            -(3-2\sqrt{2})               & 1                             & 0 \\
            -1                           & -(3-2\sqrt{2})                & A \\
            \frac{\sqrt{2}-1}{\sqrt{6}}A & -\frac{\sqrt{2}+1}{\sqrt{6}}A & 0
        \end{array}\right)\nonumber\\
        &+&\lambda^2\left(
        \begin{array}{ccc}
            -(30\sqrt{3}-21\sqrt{6})        & 0                                 & (\rho-i\eta)A \\
            \frac{\sqrt{2}-1}{2\sqrt{6}}A^2 &-(30\sqrt{3}-21\sqrt{6})-\frac{\sqrt{2}+1}{2\sqrt{6}}A^2 & 0 \\
            \left(1-\frac{\sqrt{2}+1}{\sqrt{6}}(\rho+i\eta)\right)A
            & \left(3-2\sqrt{2}-\frac{\sqrt{2}-1}{\sqrt{6}}(\rho+i\eta)\right)A & -{1\over2}A^2
        \end{array}\right)+\mathcal{O}(\lambda^3). 
\end{eqnarray}

The Jarlskog parameters for quark and lepton have been obtained in
Ref.~\cite{he} and Ref.~\cite{pakvasa}  (to lowest order of
$\epsilon^{Q,L}_{ij}$), respectively.
\begin{eqnarray}
&&{J}^Q=\left(\frac{1}{6}+\frac{2\sqrt{2}}{3}\epsilon^Q_{12}\right)
\epsilon^Q_{23}\epsilon^Q_{13}\sin\delta^Q.\nonumber\\
&&{J}^L=\left(\frac{1}{3\sqrt{2}}+\frac{1}{6}\epsilon^L_{12}\right)
\epsilon^L_{13}\sin\delta^L.
\end{eqnarray}

\section{Discussions and Conclusions}

In the literature there are several different zeroth order basis for
the CKM and PMNS matrices. Each of them has its virtual. We would
like to comment on a few of them before drawing our conclusions. We
find that although some of these parametrizations converge faster
than the ones in the previous two sections, they are in general very
complicated in expression for perturbation series and difficult to
use.

The first case is a variant of QLC. Petcov and Smirnov~\cite{petcov}
noticed that $\tan{2\theta^L_{12}}=1/\tan{2\theta^Q_{12}}$. This is
in fact another form of QLC. In Ref.~\cite{duret}, specific values
for the zeroth order angles are given with
\begin{eqnarray}
\tan{2\theta^Q_{12}}=\frac{1}{2},\quad\tan{2\theta^L_{12}}=2.\label{theta12}
\end{eqnarray}
which gives that $\sin\theta^Q_{12}=0.2298$ and agrees very well
with $\sin\theta^Q_{12}=0.2257\pm 0.0010$ obtained from
Eq.~(\ref{vv}). In this case $\epsilon^Q_{12}$ is smaller than the
cases discussed in Sections III and IV. From convergency point of
view this may be a better zeroth order basis. We have worked out the
corresponding basis $V_0$ in Appendix A. It can be seen that the
zeroth order basis is complicated which makes one wonder if it is a
good way to expand the mixing matrix. The expansion is too
complicated to be useful as can be seen from the results displayed
in Appendix A.

There are also parameterizations for quark and lepton mixing without reference to QLC.
We comment on two cases here.
One of them is to set~\cite{frampton}
\begin{eqnarray}
&&\tan{2\theta^Q_{12}}=\frac{\sqrt{2}}{3}\;,\quad\theta^Q_{23}=0\;,\quad\theta^Q_{13}=0\;,\nonumber\\
&&\tan{\theta^L_{12}}=\frac{\sqrt{2}}{2}\;,\quad\theta^L_{23}=\frac{\pi}{4}\;,\quad\theta^L_{13}=0\;.
\end{eqnarray}
The above set of parametrization violate the QLC relations since
$\theta^Q_{12}+\theta^L_{12}=47.9^\circ$. The basis matrix for the
PMNS matrix is still the tri-bimaximal mixing matrix, and the
triminimal parametrization of the PMNS matrix has already been
derived in Ref.~\cite{pakvasa}. However, as can be seen from
Appendix B, the basis matrix for quark mixing is very complicated.
One concludes again that this way of expanding the mixing matrix is
too complicated to be practical for using. The above two basis can
not be considered, at least, as convenient ones to use.

Another is the case discussed in Ref.~\cite{rodejohann} where the
basis angles are chosen to be
\begin{eqnarray}
&&\theta^Q_{12}=\frac{\pi}{12}\;,\quad\theta^Q_{23}=0\;,\quad\theta^Q_{13}=0\;,\nonumber\\
&&\theta^L_{12}=\frac{\pi}{5}\;,\quad\theta^L_{23}=\frac{\pi}{4}\;,\quad\theta^L_{13}=0\;.
\end{eqnarray}
It is interesting to note that the angle $\pi/5$ gives
$\cos(\theta^L_{12})$ to be half of the golden ratio
$(1+\sqrt{5})/2$. The basis matrices are reasonably simple.

The above two cases, however, leave the basis matrices for quarks
and leptons unrelated. If one is looking for a unified
parametrization for quark and lepton sectors, this may not be a good
way to go.

We now summarize the main results. We have derived the general expressions
of the triminimal parametrizations of the mixing matrix for quark and lepton sectors.
The triminimal parametrization is an approximate
expansion based on the standard parametrization of the mixing matrix
in three angle parameters and one phase parameter. When the zeroth order basis of
the mixing matrix is determined, the parameters of the triminimal
parametrization are fixed.

Compared with the familiar Wolfenstein-like parametrization, the
triminimal parametrization has the advantages of uniqueness and
simplicity. In Wolfenstein (-like) parametrizations, the correction
terms are put in by hand with certain physical consideration taken
into account. The explicit form has ambiguities depending on one's
preference of physical considerations even the zeroth order basis
matrices are chosen. This makes the use of the parametrization more
complicated. Of course if the parametrizations have been done
consistently, different parametrizations must be able to transfer
from one to another. We have explicitly shown how to identify the
connections of parameters in the triminimal and Wolfenstein (-like)
parametrizations for several interesting choices of basis matrices,
the unit, bimaximal and tri-bimaximal cases. The Wolfenstein-like
parametrizations derived from triminimal parametrizations have the
advantage of the uniqueness whereas as convenient to use as
traditional Wolfenstein parametrization in practical applications. 

A priori, parametrization for quark and lepton sectors are seemly
not connected. However, if connection can be made, it may provide a
hint for the underlying theory generating mixing. The QLC relations
provide a useful guide. We find that the unit basis for quark mixing
corresponds to the bimaximal basis for lepton mixing. Present
experimental data indicate that the tri-bimaximal pattern represents
mixing in the lepton sector very well. We have therefore studied the
corresponding basis matrix in the quark sector and compared with
previous studies based on Wolfenstein-like parametrizations. If one
imposes the condition that the parametrization must satisfy the QLC
relations even when corrections to the basis matrices are included,
then the corrections $\epsilon^Q_{ij}$ for quark and
$\epsilon^L_{ij}$ for lepton sectors are related by $\epsilon^Q_{12}
+ \epsilon^L_{12} = 0$ and $\epsilon^Q_{23} + \epsilon^L_{23} = 0$.
One should keep in mind that the corrections are not necessarily
required to satisfy these relations. With more accurate data to be
available in the near future, the QLC relations can be tested.

For perturbative expansions of the CKM and PMNS matrices, one should
use simple zeroth order basis matrices and the expansions should
have a fast convergence. In the literature, many other zeroth order
basis matrices have been proposed. We find some of them very
complicated to use and some of them have the quark and lepton mixing
matrix unrelated. We find that the pair simple zeroth order basis
matrices, the unit matrix for quarks and bimaximal matrix for
leptons, are convenient to use with reasonable convergency. The pair
simple zeroth order basis matrices, the tri-bimaximal matrix for
leptons and its QLC partner for quarks, make the expansions converge
faster. We consider these two sets of zeroth basis matrices as good
starting point for perturbative parametrizations for the CKM and
PMNS matrices. It would be interesting to see if theoretical
progresses can realize such connections.

\begin{acknowledgments}
This work was partially supported by NSFC (Nos.~10721063, 10575003,
10528510), by the Key Grant Project of Chinese Ministry of Education
(No.~305001), by NSC, and by NCTS.
\end{acknowledgments}

\begin{center}
{\bf APPENDIX A}
\end{center}

The basis matrices for the CKM and PMNS matrices with $\tan{2\theta^Q_{12}}=\frac{1}{2}$ and $\tan{2\theta^L_{12}}=2$ are
\begin{eqnarray}
&&V_{\rm{CKMb}}=\left(
              \begin{array}{ccc}
                (10-4\sqrt{5})^{-1/2}  & \frac{1}{2\sqrt{5}}(10-4\sqrt{5})^{1/2} & 0 \\
                -\frac{1}{2\sqrt{5}}(10-4\sqrt{5})^{1/2} & (10-4\sqrt{5})^{-1/2} & 0 \\
                0 & 0 & 1 \\
              \end{array}
            \right),\nonumber\\
&&U_{\rm{PMNSb}}=\left(
              \begin{array}{ccc}
                2(10-2\sqrt{5})^{-1/2}           & \frac{1}{2\sqrt{5}}(10-2\sqrt{5})^{1/2} & 0 \\
                -\frac{1}{2\sqrt{10}}(10-4\sqrt{5})^{1/2} & \sqrt{2}(10-2\sqrt{5})^{-1/2}  & \frac{\sqrt{2}}{2} \\
                \frac{1}{2\sqrt{10}}(10-4\sqrt{5})^{1/2}  & -\sqrt{2}(10-2\sqrt{5})^{-1/2} & \frac{\sqrt{2}}{2} \\
              \end{array}
            \right).
\end{eqnarray}
We see that the basis matrices are very complicated.

For triminimal expansion, we set
\begin{eqnarray}
&&\theta^Q_{12}=\arctan{(\sqrt{5}-2)}+\epsilon^Q_{12},
\quad\theta^Q_{23}=\epsilon^Q_{23},\quad\theta^Q_{13}=\epsilon^Q_{13},\nonumber\\
&&\theta^L_{12}=\arctan\frac{\sqrt{5}-1}{2}+\epsilon^L_{12},
\quad\theta^L_{23}=\frac{\pi}{4}+\epsilon^L_{23},\quad\theta^L_{13}=\epsilon^L_{13},
\end{eqnarray}
then we have
\begin{eqnarray}
V_{\rm{CKM}}&=&R_{23}(\epsilon^Q_{23})U^\dag_\delta
R_{13}(\epsilon^Q_{13})U_\delta
R_{12}(\epsilon^Q_{12})R_{12}(\arctan{(\sqrt{5}-2)}),\nonumber\\
U_{\rm{PMNS}}&=&R_{23}(\frac{\pi}{4})R_{23}(\epsilon^L_{23})U^\dag_\delta
R_{13}(\epsilon^L_{13})U_\delta
R_{12}(\epsilon^L_{12})R_{12}(\arctan\frac{\sqrt{5}-1}{2}).
\end{eqnarray}

To second order in $\epsilon_{ij}$, we have
\begin{eqnarray}
&&V_{\rm{CKM}}= V_{\rm{CKMb}}
+ \epsilon^Q_{12}\left(\begin{array}{ccc}  -\frac{1}{2\sqrt{5}}(10-4\sqrt{5})^{1/2} &  (10-4\sqrt{5})^{-1/2} & 0 \\
                                           -(10-4\sqrt{5})^{-1/2} & -\frac{1}{2\sqrt{5}}(10-4\sqrt{5})^{1/2} & 0 \\
                                           0 & 0 & 0 \\
                                           \end{array}
                                           \right)
                                           \nonumber\\
  &+&\epsilon^Q_{23}\left(
                  \begin{array}{ccc}
                    0 & 0 & 0 \\
                    0 & 0 & 1 \\
                    \frac{1}{2\sqrt{5}}(10-4\sqrt{5})^{1/2} & -(10-4\sqrt{5})^{-1/2} & 0 \\
                  \end{array}
                \right)
  +\epsilon^Q_{13}\left(
                  \begin{array}{ccc}
                    0 & 0 & e^{-i\delta^Q} \\
                    0 & 0 & 0 \\
                    -(10-4\sqrt{5})^{-1/2}e^{i\delta^Q} & -\frac{1}{2\sqrt{5}}(10-4\sqrt{5})^{1/2}e^{i\delta^Q} & 0 \\
                  \end{array}
                \right)\nonumber\\
  &+&(\epsilon^Q_{12})^2\left(
                  \begin{array}{ccc}
                    -\frac{1}{2}(10-4\sqrt{5})^{-1/2} & -\frac{1}{4\sqrt{5}}(10-4\sqrt{5})^{1/2} & 0 \\
                    \frac{1}{4\sqrt{5}}(10-4\sqrt{5})^{1/2} & -\frac{1}{2}(10-4\sqrt{5})^{-1/2} & 0 \\
                    0 & 0 & 0 \\
                  \end{array}
                \right)
  +(\epsilon^Q_{23})^2\left(
                  \begin{array}{ccc}
                    0 & 0 & 0 \\
                    \frac{1}{4\sqrt{5}}(10-4\sqrt{5})^{1/2} & -\frac{1}{2}(10-4\sqrt{5})^{-1/2} & 0 \\
                    0 & 0 & -\frac{1}{2} \\
                  \end{array}
                \right)
                \nonumber\\
  &+&(\epsilon^Q_{13})^2\left(
                  \begin{array}{ccc}
                    -\frac{1}{2}(10-4\sqrt{5})^{-1/2} & -\frac{1}{4\sqrt{5}}(10-4\sqrt{5})^{1/2} & 0 \\
                    0 & 0 & 0 \\
                    0 & 0 & -\frac{1}{2} \\
                  \end{array}
                \right)
 + \epsilon^Q_{12}\epsilon^Q_{23}\left(
                  \begin{array}{ccc}
                    0 & 0 & 0 \\
                    0 & 0 & 0 \\
                    (10-4\sqrt{5})^{-1/2} & \frac{1}{2\sqrt{5}}(10-4\sqrt{5})^{1/2} & 0 \\
                  \end{array}
                \right)
                \nonumber\\
  &+&\epsilon^Q_{12}\epsilon^Q_{13}e^{i\delta^Q}\left(
                  \begin{array}{ccc}
                    0 & 0 & 0 \\
                    0 & 0 & 0 \\
                    \frac{1}{2\sqrt{5}}(10-4\sqrt{5})^{1/2} & -(10-4\sqrt{5})^{-1/2} & 0 \\
                  \end{array}
                \right)
  +\epsilon^Q_{23}\epsilon^Q_{13}e^{i\delta^Q}\left(
                  \begin{array}{ccc}
                    0 & 0 & 0 \\
                    -(10-4\sqrt{5})^{-1/2} & -\frac{1}{2\sqrt{5}}(10-4\sqrt{5})^{1/2} & 0 \\
                    0 & 0 & 0 \\
                  \end{array}
                \right)\nonumber\\
  &+&\mathcal{O}\left((\epsilon^Q_{ij})^3\right).
\end{eqnarray}

\begin{eqnarray}
&&U_{\rm{PMNS}}= U_{\rm{PMNSb}}
+ \epsilon^L_{12}\left(\begin{array}{ccc}-\frac{1}{2\sqrt{5}}(10-2\sqrt{5})^{1/2} & 2(10-2\sqrt{5})^{-1/2} & 0 \\
                                           -\sqrt{2}(10-2\sqrt{5})^{-1/2}
                                           & -\frac{1}{2\sqrt{10}}(10-2\sqrt{5})^{1/2} & 0 \\
                                           \sqrt{2}(10-2\sqrt{5})^{-1/2}
                                           & \frac{1}{2\sqrt{10}}(10-2\sqrt{5})^{1/2} & 0 \\
                                           \end{array}
                                           \right)\nonumber\\
  &+&\epsilon^L_{23}\left(
                  \begin{array}{ccc}
                    0 & 0 & 0 \\
                    \frac{1}{2\sqrt{10}}(10-2\sqrt{5})^{1/2} & -\sqrt{2}(10-2\sqrt{5})^{-1/2} & \frac{\sqrt{2}}{2} \\
                    \frac{1}{2\sqrt{10}}(10-2\sqrt{5})^{1/2} & -\sqrt{2}(10-2\sqrt{5})^{-1/2} & -\frac{\sqrt{2}}{2} \\
                  \end{array}
                \right)\nonumber\\
  &+&\epsilon^L_{13}\left(
                  \begin{array}{ccc}
                    0 & 0 & e^{-i\delta^L} \\
                    -\sqrt{2}(10-2\sqrt{5})^{-1/2}e^{i\delta^L}
                    & -\frac{1}{2\sqrt{10}}(10-2\sqrt{5})^{1/2}e^{i\delta^L} & 0 \\
                    -\sqrt{2}(10-2\sqrt{5})^{-1/2}e^{i\delta^L}
                    & -\frac{1}{2\sqrt{10}}(10-2\sqrt{5})^{1/2}e^{i\delta^L} & 0 \\
                  \end{array}
                \right)\nonumber\\
  &+&(\epsilon^L_{12})^2\left(
                  \begin{array}{ccc}
                    -(10-2\sqrt{5})^{-1/2} & -\frac{1}{4\sqrt{5}}(10-2\sqrt{5})^{1/2} & 0 \\
                    \frac{1}{4\sqrt{10}}(10-2\sqrt{5})^{1/2} & -\frac{1}{\sqrt{2}}(10-2\sqrt{5})^{-1/2} & 0 \\
                    -\frac{1}{4\sqrt{10}}(10-2\sqrt{5})^{1/2} & \frac{1}{\sqrt{2}}(10-2\sqrt{5})^{-1/2} & 0 \\
                  \end{array}
                \right)
  +(\epsilon^L_{23})^2\left(
                  \begin{array}{ccc}
                    0 & 0 & 0 \\
                    \frac{1}{4\sqrt{10}}(10-2\sqrt{5})^{1/2}
                    & -\frac{1}{\sqrt{2}}(10-2\sqrt{5})^{-1/2} & -\frac{\sqrt{2}}{4} \\
                    -\frac{1}{4\sqrt{10}}(10-2\sqrt{5})^{1/2}
                    & \frac{1}{\sqrt{2}}(10-2\sqrt{5})^{-1/2} & -\frac{\sqrt{2}}{4} \\
                  \end{array}
                \right)
                \nonumber\\
  &+&(\epsilon^L_{13})^2\left(
                  \begin{array}{ccc}
                    -(10-2\sqrt{5})^{-1/2} & -\frac{1}{4\sqrt{5}}(10-2\sqrt{5})^{1/2} & 0 \\
                    0 & 0 & -\frac{\sqrt{2}}{4} \\
                    0 & 0 & -\frac{\sqrt{2}}{4} \\
                  \end{array}
                \right)
 + \epsilon^L_{12}\epsilon^L_{23}\left(
                  \begin{array}{ccc}
                    0 & 0 & 0 \\
                    \sqrt{2}(10-2\sqrt{5})^{-1/2} & \frac{1}{2\sqrt{10}}(10-2\sqrt{5})^{1/2} & 0 \\
                    \sqrt{2}(10-2\sqrt{5})^{-1/2} & \frac{1}{2\sqrt{10}}(10-2\sqrt{5})^{1/2} & 0 \\
                  \end{array}
                \right)
                \nonumber\\
  &+&\epsilon^L_{12}\epsilon^L_{13}e^{i\delta^L}\left(
                  \begin{array}{ccc}
                    0 & 0 & 0 \\
                    \frac{1}{2\sqrt{10}}(10-2\sqrt{5})^{1/2} & -\sqrt{2}(10-2\sqrt{5})^{-1/2} & 0 \\
                    \frac{1}{2\sqrt{10}}(10-2\sqrt{5})^{1/2} & -\sqrt{2}(10-2\sqrt{5})^{-1/2} & 0 \\
                  \end{array}
                \right)\nonumber\\
  &+&\epsilon^L_{23}\epsilon^L_{13}e^{i\delta^L}\left(
                  \begin{array}{ccc}
                    0 & 0 & 0 \\
                    -\sqrt{2}(10-2\sqrt{5})^{-1/2} & -\frac{1}{2\sqrt{10}}(10-2\sqrt{5})^{1/2} & 0 \\
                    \sqrt{2}(10-2\sqrt{5})^{-1/2}  & \frac{1}{2\sqrt{10}}(10-2\sqrt{5})^{1/2}  & 0 \\
                  \end{array}
                \right)
  +\mathcal{O}\left((\epsilon^L_{ij})^3\right).
\end{eqnarray}

\begin{center}
{\bf APPENDIX B}
\end{center}

With the following angles for the zeroth order expansion
\begin{eqnarray}
&&\tan{2\theta^Q_{12}}=\frac{\sqrt{2}}{3},\quad\theta^Q_{23}=0,\quad\theta^Q_{13}=0,\nonumber\\
&&\tan{\theta^L_{12}}=\frac{\sqrt{2}}{2},\quad\theta^L_{23}=\frac{\pi}{4},\quad\theta^L_{13}=0.
\end{eqnarray}
we obtain the basis matrix for the CKM matrix
\begin{eqnarray}
&&V_{\rm{CKMb}}=\left(
              \begin{array}{ccc}
                \frac{1}{11^{1/4}}(\sqrt{11}-3)^{-1/2}  & \frac{1}{11^{1/4}\sqrt{2}}(\sqrt{11}-3)^{1/2} & 0 \\
                -\frac{1}{11^{1/4}\sqrt{2}}(\sqrt{11}-3)^{1/2} & \frac{1}{11^{1/4}}(\sqrt{11}-3)^{-1/2} & 0 \\
                0 & 0 & 1 \\
              \end{array}
            \right),
\end{eqnarray}
and
\begin{eqnarray}
V_{\rm{CKM}}&=&R_{23}(\epsilon^Q_{23})U^\dag_\delta
R_{13}(\epsilon^Q_{13})U_\delta
R_{12}(\epsilon^Q_{12})R_{12}(\arctan{\frac{\sqrt{11}-3}{\sqrt{2}}}).
\end{eqnarray}
The triminimal parametrization of the CKM matrix is
\begin{eqnarray}
&&V_{\rm{CKM}}\nonumber\\ &=&V_{\rm{CKMb}}
+ \frac{\epsilon^Q_{12}}{11^{1/4}}\left(\begin{array}{ccc}-\frac{1}{\sqrt{2}}(\sqrt{11}-3)^{1/2}
                                           & (\sqrt{11}-3)^{-1/2} & 0 \\
                                           -(\sqrt{11}-3)^{-1/2} & -\frac{1}{\sqrt{2}}(\sqrt{11}-3)^{1/2} & 0 \\
                                           0 & 0 & 0 \\
                                           \end{array}
                                           \right)
  +\frac{\epsilon^Q_{23}}{11^{1/4}}\left(
                  \begin{array}{ccc}
                    0 & 0 & 0 \\
                    0 & 0 & 11^{1/4} \\
                    \frac{1}{\sqrt{2}}(\sqrt{11}-3)^{1/2} & -(\sqrt{11}-3)^{-1/2} & 0 \\
                  \end{array}
                \right)\nonumber\\
  &+&\frac{\epsilon^Q_{13}}{11^{1/4}}\left(
                  \begin{array}{ccc}
                    0 & 0 & 11^{1/4}e^{-i\delta^Q} \\
                    0 & 0 & 0 \\
                    -(\sqrt{11}-3)^{-1/2}e^{i\delta^Q} & -\frac{1}{\sqrt{2}}(\sqrt{11}-3)^{1/2}e^{i\delta^Q} & 0 \\
                  \end{array}
                \right)\nonumber\\
  &+&\frac{(\epsilon^Q_{12})^2}{11^{1/4}}\left(
                  \begin{array}{ccc}
                    -\frac{1}{2}(\sqrt{11}-3)^{-1/2} & -\frac{1}{2\sqrt{2}}(\sqrt{11}-3)^{1/2} & 0 \\
                    \frac{1}{2\sqrt{2}}(\sqrt{11}-3)^{1/2} & -\frac{1}{2}(\sqrt{11}-3)^{-1/2} & 0 \\
                    0 & 0 & 0 \\
                  \end{array}
                \right)
  +\frac{(\epsilon^Q_{23})^2}{11^{1/4}}\left(
                  \begin{array}{ccc}
                    0 & 0 & 0 \\
                    \frac{1}{2\sqrt{2}}(\sqrt{11}-3)^{1/2} & -\frac{1}{2}(\sqrt{11}-3)^{-1/2} & 0 \\
                    0 & 0 & -\frac{11^{1/4}}{2} \\
                  \end{array}
                \right)
                \nonumber\\
  &+&\frac{(\epsilon^Q_{13})^2}{11^{1/4}}\left(
                  \begin{array}{ccc}
                    -\frac{1}{2}(\sqrt{11}-3)^{-1/2} & -\frac{1}{2\sqrt{2}}(\sqrt{11}-3)^{1/2} & 0 \\
                    0 & 0 & 0 \\
                    0 & 0 & -\frac{11^{1/4}}{2} \\
                  \end{array}
                \right)
 + \frac{\epsilon^Q_{12}\epsilon^Q_{23}}{11^{1/4}}\left(
                  \begin{array}{ccc}
                    0 & 0 & 0 \\
                    0 & 0 & 0 \\
                    (\sqrt{11}-3)^{-1/2} & \frac{1}{\sqrt{2}}(\sqrt{11}-3)^{1/2} & 0 \\
                  \end{array}
                \right)
                \nonumber\\
  &+&\frac{\epsilon^Q_{12}\epsilon^Q_{13}e^{i\delta^Q}}{11^{1/4}}\left(
                  \begin{array}{ccc}
                    0 & 0 & 0 \\
                    0 & 0 & 0 \\
                    \frac{1}{\sqrt{2}}(\sqrt{11}-3)^{1/2} & -(\sqrt{11}-3)^{-1/2} & 0 \\
                  \end{array}
                \right)
  +\frac{\epsilon^Q_{23}\epsilon^Q_{13}e^{i\delta^Q}}{11^{1/4}}\left(
                  \begin{array}{ccc}
                    0 & 0 & 0 \\
                    -(\sqrt{11}-3)^{-1/2} & -\frac{1}{\sqrt{2}}(\sqrt{11}-3)^{1/2} & 0 \\
                    0 & 0 & 0 \\
                  \end{array}
                \right)\nonumber\\
  &+&\mathcal{O}\left((\epsilon^Q_{ij})^3\right).
\end{eqnarray}

\end{document}